\documentclass[lettersize,journal]{IEEEtran}
\usepackage{amsmath,amsfonts}
\usepackage{algorithmic}
\usepackage{algorithm}
\usepackage{array}
\usepackage{textcomp}
\usepackage{stfloats}
\usepackage{url}
\usepackage{verbatim}
\usepackage{graphicx}
\usepackage{cite}
\usepackage{booktabs}
\usepackage{multirow}
\usepackage{subfigure}
\usepackage{wrapfig}
\usepackage{hyperref}
\usepackage{color}

\begin{document}

\title{Rate-Distortion Optimized Post-Training Quantization for Learned Image Compression}

\author{Junqi Shi, Ming Lu,~\IEEEmembership{Member,~IEEE}, and Zhan Ma,~\IEEEmembership{Senior Member,~IEEE}

\thanks{This work was partially supported by the National Natural Science Foundation of China under Grant 62022038.}
\thanks{Copyright © 20xx IEEE. Personal use of this material is permitted. However, permission to use this material for any other purposes must be obtained from the IEEE by sending an email to pubs-permissions@ieee.org.}
\thanks{Authors are with the School of Electronic Science and Engineering, Nanjing University, Nanjing 210093, P.R. China. Emails: junqishi@smail.nju.edu.cn, minglu@nju.edu.cn, mazhan@nju.edu.cn. {\it J. Shi and M. Lu contributed equally to this work.} (Corresponding Author: Z. Ma)}
}

\maketitle
\begin{abstract}
Quantizing a floating-point neural network to its fixed-point representation is crucial for Learned Image Compression (LIC) because {it improves decoding consistency for interoperability and reduces space-time complexity for implementation.} Existing solutions often have to retrain the network for model quantization, which is time-consuming and impractical to some extent. This work suggests using Post-Training Quantization (PTQ) to process pretrained, off-the-shelf LIC models. We theoretically prove that minimizing quantization-induced mean square error (MSE) of model parameters (e.g., weight, bias, and activation) in PTQ is sub-optimal for compression tasks and thus develop a novel Rate-Distortion (R-D) Optimized PTQ (RDO-PTQ) to best retain the compression performance. Given a  LIC model,  RDO-PTQ  layer-wisely determines the quantization parameters to transform the original floating-point parameters in 32-bit precision (FP32) to fixed-point ones at 8-bit precision (INT8), for which a tiny calibration image set is compressed in optimization to minimize R-D loss. Experiments reveal the outstanding efficiency of the proposed method on different LICs, showing the closest coding performance to their floating-point counterparts. Our method is a lightweight and plug-and-play approach {without retraining model parameters but just adjusting quantization parameters,} which is attractive to practitioners. Such an RDO-PTQ is a task-oriented PTQ scheme, which is then extended to quantize popular super-resolution and image classification models with negligible performance loss, further evidencing the generalization of our methodology. Related materials will be released at \url{https://njuvision.github.io/RDO-PTQ}.
\end{abstract}


\section{Introduction}
\IEEEPARstart{C}{ompressed} images are vastly used in networked applications for efficient information sharing, which has continuously driven the pursuit of better compression technologies for the past decades \cite{JPEG,BPG,VVC}. Built upon the advances of deep neural networks (DNNs), recent years have witnessed the explosive growth of learned image compression solutions \cite{ballemshj18,minnenbt18,chen2021end,cheng2020learned,hu2021learning,TIC,guo2021causal} that offer superior efficiency to conventional, well-known rules-based JPEG \cite{JPEG}, HEVC Intra (BPG) \cite{BPG}, and even Versatile Video Coding Based Intra Profile (VVC Intra) \cite{VVC}. 

\begin{figure}
    \centering
    \includegraphics[width=0.95\columnwidth]{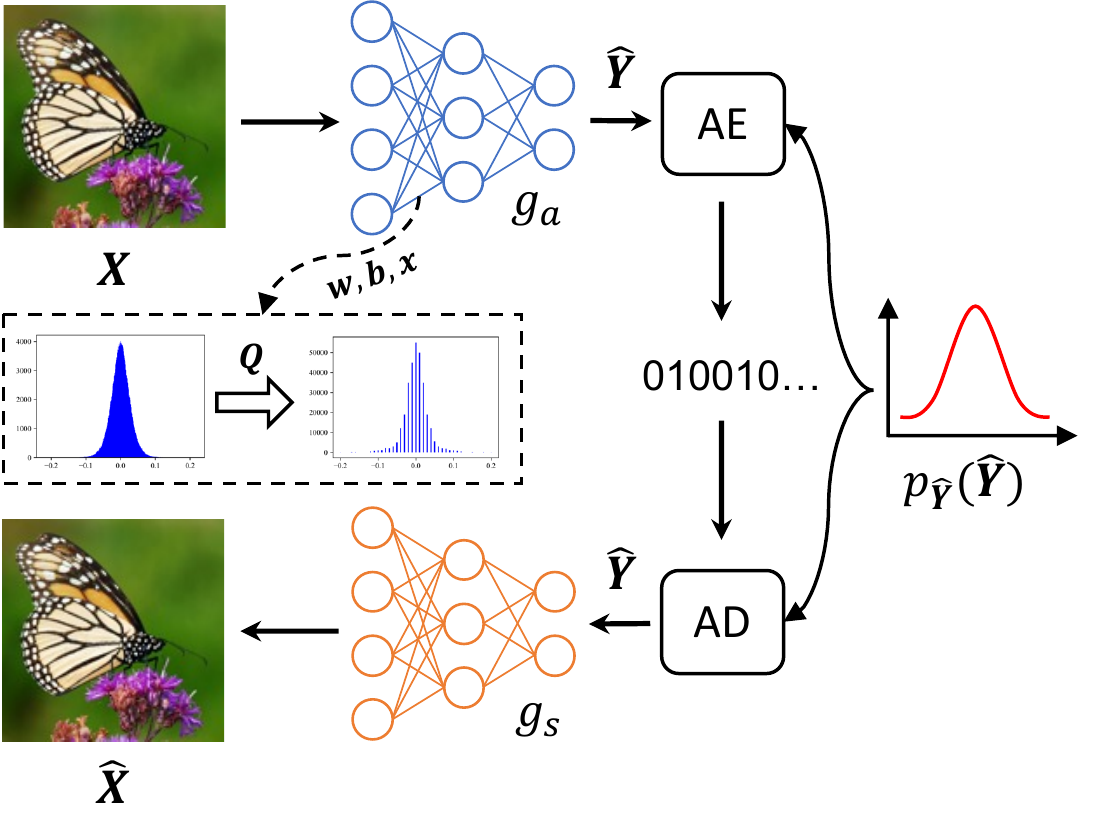}
    \caption{{\it Learned Image Compression (LIC).} $g_a$ ($g_s$) is main encoder (decoder); AE/AD is arithmetic encoding/decoding using $p_{\hat{\boldsymbol{Y}}}(\hat{\boldsymbol{Y}})$. Either convolution or self-attention is used to derive  $\hat{\boldsymbol{Y}}$ from input $\boldsymbol{X}$. Quantization $\boldsymbol{Q}$ is applied at every layer (convolutional or self-attention) to transform weight $\boldsymbol{w}$, bias $\boldsymbol{b}$, and activation $\boldsymbol{x}$ from their native FP32 precision to INT8 format.}
    \label{fig0}
\end{figure}

\subsection{Background \& Motivation} 
Nevertheless, existing learned image compression (LIC) approaches typically adopt the floating-point format for data representation (e.g., weight, bias, activation), which not only consumes an excessive amount of space-time complexity but also leads to platform inconsistency and decoding failures \cite{he2022PTQ}. Numerous explorations have been made to tackle these for practical applications, {in which model pruning and quantization are two promising directions under consideration}. Model pruning \cite{molchanov2019importance, guo2020model} is to trim redundant parameters while retaining the model's capacity. After pruning, the space-time complexity and platform inconsistency still persist because the pruned model remains in floating-point format. Additionally, pruning is not consistently effective due to the varying redundancy of different models. {On the other hand, model quantization is commonly employed to generate fixed-point (or integer) LICs \cite{balle2018integer, hong2020efficient, sun2021learned}, aiming to reduce complexity and enhance platform consistency. Quantization techniques can be further categorized as follows.}

{Classical Quantization-Aware Training (QAT), as discussed in \cite{bhalgat2020lsq+, le2022mobilecodec, sun2021learned, xu2022improving}, has primarily been employed in \cite{balle2018integer, hong2020efficient, sun2020end, sun2021learned} to transform floating-point LICs into their fixed-point counterparts. However, these methods necessitate retraining model parameters with complete access to labels, which can be costly and impractical.}

Recently, Post-Training Quantization (PTQ) \cite{nagel2020up,nagel2021white} offered a push-button solution to directly quantize pretrained, off-the-shelf network models without retraining model parameters. However, existing PTQ methods were mainly dedicated to high-level vision tasks \cite{Low-Bit-PTQ,liu2021post, li2020brecq, nagel2020up}. {To date, there has been limited development of PTQ schemes tailored for quantizing LIC~\cite{sun2022q}. This work, therefore, extends the application of PTQ to image compression model quantization.}

\subsection{Approach \& Contribution} 
{In the context of image compression, the task loss is measured by considering both the rate and distortion jointly rather than relying solely on distortion measurements. Our theoretical analysis establishes that exclusively minimizing quantization distortion (e.g., MSE) of model parameters, a common practice in traditional PTQ methods \cite{Low-Bit-PTQ,liu2021post}, is sub-optimal from the compression perspective. This suboptimality arises due to the localized non-monotonic relationship between quantization error and rate-distortion performance in the compression task\footnote{We also experimentally show that explicitly optimizing R-D loss outperforms methods like AdaRound and BRECQ, which implicitly model the task loss through per-layer second-order approximations~\cite{li2020brecq,nagel2020up}.}.} 

We then propose the Rate-Distortion Optimized PTQ (RDO-PTQ) to jointly consider the reconstruction distortion and compression bitrate (e.g., bpp - bits per pixel) when optimizing the quantization parameters to transform 32-bit floating-point (FP32) parameters to corresponding 8-bit fixed-point (INT8) representation for any pretrained, off-the-shelf LIC model.


Considering the optimization complexity and generalization, the RDO-PTQ is executed from one network layer\footnote{For simplicity, we refer to the ``network layer'' as the ``layer''. } to another (e.g., layer-wise) to {determine proper model quantization parameters (e.g., range and offset) of weight, bias, and activation in convolutional or self-attention units.} The current implementation compresses a tiny calibration image set (e.g., less than ten images) to decide model quantization parameters.

{As the distribution of both weight and activation varies across channels at each layer, the range is further adapted channel-wisely on top of the layer-wise adaptation (see Fig.~\ref{eq:channel_wise_weight}).} In addition to the range determination of the bias, a bias rescaling is applied to enforce the INT8 computation strictly.


In summary, the contribution of our method is mainly reflected in the following aspects:
\begin{enumerate}

    \item {We propose a task-oriented PTQ to quantize the LIC model with the joint consideration of rate and distortion metrics, referred to as RDO-PTQ. This lightweight solution requires merely compressing a few random image samples to adjust quantization parameters without retraining model parameters.\footnote{It suggests that we directly transform a pretrained, off-the-shelf, floating-point model to its fixed-point representation for inference. Here, model parameters retraining refers to optimizing the (floating-point) weight and bias using (potentially) large-scale data samples. In contrast, our method only needs to optimize quantization parameters (e.g., scaling factors).}}
    \item Such an RDO-PTQ is robust and effective, demonstrating the closest compression efficiency between native FP32 and the corresponding quantized INT8 model of various popular LIC solutions.
    \item The proposed task-oriented PTQ is also applied to super-resolution and image classification tasks with negligible performance loss, further revealing the generalization of the methodology.
\end{enumerate}

The remaining parts of this paper are structured as follows: Sec.~\ref{sec:Related_Work} briefly reviews related works, while Sec.~\ref{sec:quantization_basics} introduces the fundamental ideas used for model quantization. The proposed RDO-PTQ is explained in detail in Sec.~\ref{sec:RDO-PTQ} followed by extensive comparisons in Sec.~\ref{sec:exp}. Conclusions are finally drawn in Sec.~\ref{sec:conclusion}. To assist the understanding, Table~\ref{tab:notations} lists notations used frequently in this paper.

\begin{table}[t]
    \centering
    \caption{Notations}
    \label{tab:notations}
    \begin{tabular}{c|c}
    \hline
       Abbreviation  & Description \\
       \hline
       LIC & Learned Image Compression\\
       QAT & Quantization-Aware Training\\
       PTQ & Post-Training Quantization\\
       RDO-PTQ & Rate-Distortion Optimized PTQ\\
       MSE & Mean Square Error\\
       FP32 & 32-bit Floating-point format\\
       INT8/INT10 & 8-bit/10-bit Fixed-point format\\
       VAE & Variational Auto-Encoder\\
         \hline
    \end{tabular}
\end{table}

\section{Related Work} \label{sec:Related_Work}

The compression and acceleration of LIC models have been a long-standing topic ever since the early stages of LIC development. In this section, we first review the development of the LIC briefly and then introduce LIC quantization and novel PTQ technologies, respectively. 

\subsection{Learned Image Compression (LIC)}

Back to 2017, Ball{\'e} et al.~\cite{balle2016end} and Chen et al.~\cite{chen2017deepcoder} both showed that stacked convolutional layers could replace the traditional transforms to form an end-to-end trainable LIC with better performance than the JPEG \cite{JPEG}. Since then, the development of LIC has grown rapidly. As shown in Fig.~\ref{fig0}, prevalent LICs are mainly built upon the Variational Auto-Encoder (VAE) architecture to find a rate-distortion optimized compact representation of input images. In \cite{ballemshj18}, besides the GDN (Generalized Divisive Normalization) based nonlinear transform, a hyperprior modeled by a factorized distribution was introduced to better capture the distribution of quantized latent features in entropy coding.  Shortly, the use of joint hyperprior and autoregressive neighbors for entropy context modeling was developed in \cite{minnenbt18}, demonstrating better efficiency than the BPG -- an HEVC Intra implementation. 

Later, stacked convolution with simple ReLU units was used in \cite{cheng2020learned,chen2021end} to replace GDN.  Additionally, the attention mechanism was augmented for better information embedding, which, for the first time, outperformed the VVC Intra. However, such an entropy context model conditioned on joint hyperprior and autoregressive neighbors primarily utilized local correlations. {Recently, Qian et al.\cite{qian2020learning}, and Kim et al.\cite{kim2022joint} expanded the utilization of correlations beyond the local context, incorporating both global and local correlations through the inclusion of additional global references.}

Recalling that the principle behind the image coding is to find content-dependent models (e.g., transform, statistical distribution) for more compact representations, apparently, solutions simply stacking convolutions struggle to efficiently characterize the content dynamics because of the fixed receptive field and fixed network parameters in a trained convolutional neural network (CNN).
In the past years, as the wide spread of self-attention-based Vision Transformer (ViT) in various tasks \cite{TIC, TinyLIC, liu2021post, lin2022fqvit}, numerous works \cite{zhu2021transformer, qian2021entroformer, TIC, TinyLIC} attempted to extend the self-attention mechanism in LIC for content-adaptive embedding. Zhu et al.~\cite{zhu2021transformer} realized nonlinear transform using Swin Transform~\cite{liu2021swin} instead of stacked convolutions, while Qian et al.~\cite{qian2021entroformer} retained CNN transform but replaced the convolution-based entropy context module with the Transformer. As extensively studied in \cite{TIC, TinyLIC}, the integrated convolution and self-attention could provide performance improvements to the VVC intra on various datasets and reduce the space-time complexity significantly. 

As shown subsequently, three popular and publicly accessible LICs are used to exemplify the efficiency of the proposed RDO-PTQ, including the Minnen2018~\cite{minnenbt18}, Cheng2020~\cite{cheng2020learned} and Lu2022~\cite{TIC}.

\subsection{LIC Model Quantization}

Although learning-based solutions remarkably improved the performance for various tasks, their native floating-point representation incurred serial problems for enabling real-world services, including excessive space-time complexity consumption, nondeterministic inconsistency across heterogeneous platforms, etc. {These issues are particularly pronounced in the context of image compression, as guaranteeing interoperability across various devices is a crucial function of a LIC solution. It has been reported that even a minor numerical rounding-off error in floating-point computations can potentially result in decoding failures or inaccurate reconstructions~\cite{balle2018integer}.}


A pioneer exploration was made in \cite{balle2018integer}  to train an integer LIC to resolve platform inconsistency and decoding failures. Subsequently, the quantization of convolutional weights was specifically treated in \cite{sun2020end, sun2021learned}. Sun et al.~\cite{sun2020end} proposed a weight clipping method to reduce the weight quantization error, and its improved version with a layer-by-layer weight finetuning was presented in \cite{sun2021learned}. Though LIC's weight quantization reported negligible coding loss~\cite{sun2020end,sun2021learned}, the activation quantization was not considered. At the same time, Hong et al. \cite{hong2020efficient} proposed layer-wise range-adaptive quantization (RAQ) for both weight and bias parameters and linear scaling for feature activation.   Le et al.~\cite{le2022mobilecodec} extended the quantization to a Transformer-based neural video codec and demonstrated the real-time decoding capabilities on a mobile device. Note that these methods mainly belong to the QAT category, requiring full access to labels for model parameter retraining, which is impractical and expensive to some extent \cite{nagel2021white}.

Recently, PTQ has attracted intensive attention \cite{nagel2021white} because it is a push-button approach without model parameters retraining, which is presumably applicable to any off-the-shelf, pretrained neural network. Unfortunately, a majority of PTQ studies were still devoted to high-level vision tasks \cite{Low-Bit-PTQ,nagel2020up,liu2021post, li2020brecq}, and there were only a few works \cite{he2022PTQ, sun2022q} attempting to apply the PTQ to the compression task. For example, He et al.~\cite{he2022PTQ} adopted a standard PTQ technique with a deterministic entropy coding scheme to eliminate cross-platform inconsistencies. A Q-LIC was recently published in~\cite{sun2022q}, in which it first studied the impact of quantization error across channels and then applied channel splitting accordingly to minimize the overall distortion.

\subsection{Optimization with PTQ}
PTQ is applied to a model after it is trained. Therefore, it avoids model parameter retraining, making it much easier for deployment in practice. However, naively quantizing a full-precision floating-point model without any optimization usually incurs significant accuracy degradation. Earlier, several works tried to reduce quantization error by minimizing the MSE between the quantized and full-precision tensor \cite{hubara2020improving}. 

In the past years, an increasing number of explorations~\cite{nagel2020up, hubara2020improving, li2020brecq} realized that minimizing quantization-induced MSE  may lead to a sub-optimal solution, as the MSE does not equivalently reflect the task loss. Therefore, the task loss that directly connects with the model objectives, like classification accuracy, was extensively studied. For example, both AdaRound \cite{nagel2020up} and BRECQ \cite{li2020brecq} modeled the task loss using per-layer second-order approximation, upon which AdaRound applied adaptive rounding instead of simple rounding-to-nearest, and BRECQ further considered the cross-layer dependency by optimizing the block reconstruction.  These task loss optimized PTQ approaches focused on weight quantization, while activation quantization was largely overlooked. In the meantime,  high-level vision tasks considered in these works are constrained mainly by a single loss metric like the accuracy measurement in classification or segmentation. In contrast, the compression task cares about the rate and distortion jointly.

Given the advantages of PTQ, this work extends it to the image compression task. Since image compression pursues optimal rate-distortion performance, a novel rate-distortion optimized PTQ is developed to fulfill this purpose.

\section{Quantization Fundamental} \label{sec:quantization_basics}

This section introduces the fundamental rules of the PTQ scheme that we will use in our method. This is so-called uniform quantization, and it is widely adopted because it permits efficient implementation of fixed-point arithmetic. 

For an arbitrary floating-point value $x_{float}$ (e.g., weight, bias, or activation in either a convolutional or self-attention layer), it is quantized using:
\begin{equation}
    x_{int} = clip(\lfloor \frac{x_{float}}{s_x} \rceil + z_x, 0, 2^{b}-1), \label{1}
\end{equation}
where $\lfloor \cdot \rceil$ rounds the input to its corresponding integer. $b$ is the bit-width. $z_x$ is the zero-point offset measuring the distance shifted from the center of the quantization range, and $s_x$ is the linear scaling factor.

{Following \eqref{1}, an input $x_{float}$ is converted to the integer $x_{int}$ in the range of $[ 0, 2^{b}-1 ]$.} Then $x_{int}$ is mapped to a fixed-point $\hat{x}$ for inference \cite{hong2020efficient} or simulating the effect of integer quantization to avoid gradient vanishing  \cite{dai2021vs} in training.  Thus, the mapping is often formulated as 
\begin{equation}
    \hat{x} = s_x \cdot (x_{int} - z_x). \label{2}
\end{equation} Such mapping is also used in our RDO-PTQ for optimizing quantization parameters by examining the compression performance of a tiny set of images in the task inference stage to derive the corresponding fixed-point model.

Usually, zero-point $z_x$ is determined by the minimum values of $x_{float}$,
\begin{equation}
    z_x = clip(\lfloor \frac{\min(x_{float})}{s_x} \rceil). \label{zero-point}
\end{equation}
 $s_x$ is deduced using 
\begin{equation}
    s_x = \Gamma(\frac{r_x}{2^{b}-1}; n_r, b^s), \label{3}
\end{equation}
where $r_x$ represents the dynamic range of $x$, $b^s$ is the predefined bit-width precision of $s_x$, and $n_r$ is the reserved number of decimal digits. $\Gamma(\cdot)$ is used to generate fixed-point format, which is defined below 
\begin{equation}
    \Gamma(s;n_r,b^s) = clip(s\cdot2^{n_r}, 0, 2^{b^s}-1) \times 2^{-n_r}. \label{4}
\end{equation} In practice, $\Gamma(\cdot)$ left shifts input $s$ and cut off the extra bits to clip the value within a predefined range, and then right shift $n_r$ bits to keep the output the same order of magnitude as the input. 

\section{Rate-Distortion Optimized PTQ} \label{sec:RDO-PTQ}

In this section, we first explore the relationship between the R-D loss of LIC and quantization error and then exemplify the localized non-monotonic behavior. {Examination suggests a  smaller quantization error does not necessarily lead to a smaller R-D loss.} With this motivation, Rate-Distortion Optimized PTQ is proposed for LIC model quantization.  

\subsection{From Quantization Error to R-D Loss}
The scaling factor $s_x$ is used to map the floating-point elements to the corresponding fixed-point ones, which directly impacts the performance of the quantized fixed-point model. Thus, how to select the proper $s_x$ is crucial. To derive it as in Eq.~(\ref{3}), existing works mostly minimize the square error between vectorized elements $\boldsymbol{x}$ in FP32 and quantized  $\boldsymbol{\hat{x}}$ in INT8 (we refer to this method as ``MSE" optimization subsequently), e.g.,
\begin{equation}
    s_{\boldsymbol{x}} = {\mathop {\arg \min}_{s_{\boldsymbol{x}}}} \ \lVert \boldsymbol{x} -\hat{\boldsymbol{x}} \rVert^2. \label{eq:MMSE}
\end{equation} The use of Eq.~(\ref{eq:MMSE}) generally assumes the monotonic relation between quantization error-induced distortion $D$ (e.g., MSE) and compression efficiency measured by the R-D metric $J = R + \lambda D$~\cite{Rate-distortion-Theory}. {Apparently, such an assumption does not hold because of the inevitable rate contribution to $J$.
Besides, the MSE loss used in (\ref{eq:MMSE}) penalizes all the quantization errors equally \cite{hubara2020improving}, while in reality, the loss should penalize more quantization errors that affect the final R-D performance.}

{\bf Theoretical Justification.} 
In this section, $J(\boldsymbol{x}, \boldsymbol{w})$ represents a deep neural network that models the R-D measures between the input and output tensor, given the model parameters, e.g., weight $\boldsymbol{x}$ and activation $\boldsymbol{w}$. All network layers, including nonlinear activation functions, are defined within $J$ to process the input. For such a trained compression task $J(\boldsymbol{x},\boldsymbol{w})$ with floating-point activation $\boldsymbol{x}$ and weight $\boldsymbol{w}$, its quantized fixed-point model used for inference can be formulated as  $J(\boldsymbol{x}+\Delta{\boldsymbol{x}}, \boldsymbol{w}+\Delta{\boldsymbol{w}})$. As in other works \cite{nagel2021white, nahshan2021loss, li2020brecq}, we treat quantization noise as a perturbation with additive property, i.e., $\boldsymbol{w}+\Delta{\boldsymbol{w}}$ and $\boldsymbol{x}+\Delta{\boldsymbol{x}}$. To simplify the deduction, the bias term is neglected since it can be absorbed into $\boldsymbol{w}$ by appending a unit term to $\boldsymbol{x}$, as studied in~\cite{botev2017practical}.

Then, the quantization induced performance loss is:
\begin{align}
    \Delta{J} &= E\left[J(\boldsymbol{x}+\Delta{\boldsymbol{x}}, \boldsymbol{w}+\Delta{\boldsymbol{w}}) - J(\boldsymbol{x}, \boldsymbol{w}) \right]  \nonumber\\
                &\approx  E \left[ 
                    \nabla J \cdot \left[\begin{array}{cc}
                     \Delta \boldsymbol{x} \\
                     \Delta \boldsymbol{w} 
                    \end{array} \right] 
                    + 
                    \frac{1}{2} \left[ \begin{array}{cc}
                    \Delta \boldsymbol{x} \\
                    \Delta \boldsymbol{w}
                    \end{array} \right]^T
                    \nabla^2 {J}
                    \left [\begin{array}{cc}
                     \Delta \boldsymbol{x} \\
                     \Delta \boldsymbol{w} 
                    \end{array} \right] \right] \nonumber\\
                & \approx  E \left[ \left [\begin{array}{cc}
                    \dfrac{\partial J}{\partial \boldsymbol{x}} & \dfrac{\partial J}{\partial \boldsymbol{w}} 
                \end{array}\right ] \left [\begin{array}{cc}
                     \Delta \boldsymbol{x} \\
                     \Delta \boldsymbol{w} 
                \end{array} \right] \right] \nonumber\\
        & + \frac{1}{2} E \left[ \left[ \begin{array}{cc}
                    \Delta \boldsymbol{x} \\
                    \Delta \boldsymbol{w}
                \end{array} \right]^T
                H_{\boldsymbol{x},\boldsymbol{w}} 
                \left [\begin{array}{cc}
                     \Delta \boldsymbol{x} \\
                     \Delta \boldsymbol{w} 
                \end{array} \right] \right], 
\end{align} where $H_{\boldsymbol{x},\boldsymbol{w}}$ is the Hessian matrix, and high-order terms are ignored for simple derivation. For a converged model, its expected gradient is close to $0$, e.g., [$\tfrac{\partial J}{\partial \boldsymbol{x}} \ \tfrac{\partial J}{\partial \boldsymbol{w}}$] = 0, yielding
\begin{equation}
    \Delta J \approx \frac{1}{2} E\left[ \left[ \begin{array}{cc}
                    \Delta \boldsymbol{x} \\
                    \Delta \boldsymbol{w}
                \end{array} \right]^T \cdot H_{\boldsymbol{x}, \boldsymbol{w}} \cdot
                \left [\begin{array}{cc}
                    \Delta \boldsymbol{x} \\
                    \Delta \boldsymbol{w} 
                \end{array} \right] \right]. \label{eq:simp_task_loss}
\end{equation}

{The following example will illustrate how the minimum quantization error may not result in optimal R-D loss.} For the sake of understanding, we assume that both $\Delta \boldsymbol{x}$ and $\Delta \boldsymbol{w}$ are made up of one element. Subsequently, Eq.~(\ref{eq:simp_task_loss}) can be further expanded as:
\begin{align}
    \Delta J \approx \mbox{~}\frac{1}{2} E[\frac{\partial ^2 J}{\partial \boldsymbol{x}^2} \Delta \boldsymbol{x}^2 + \frac{2\partial ^2 J}{\partial \boldsymbol{x} \partial \boldsymbol{w}} \Delta \boldsymbol{x} \Delta \boldsymbol{w} 
    + \frac{\partial ^2 J}{\partial \boldsymbol{w}^2} \Delta \boldsymbol{w}^2]. \label{eq:expanded_simp_task_loss}
\end{align} As seen, the R-D loss not only depends on the quantization error, e.g., $\Delta \boldsymbol{x}^2$ and $\Delta \boldsymbol{w}^2$ but also is related to the second-order derivatives of $J$. Particularly, even having all positive second-order derivatives, the different signs of $\Delta \boldsymbol{x}$ and $\Delta \boldsymbol{w}$  would lead to a negative cross term $\Delta \boldsymbol{w} \Delta \boldsymbol{x}$, suggesting that a larger absolute error of quantization may not lead to a larger R-D loss.

As a toy example, suppose a symmetric Hessian matrix as
\begin{equation}
    H_{\boldsymbol{x},\boldsymbol{w}} = \left[ \begin{array}{cc}
        1 & 0.5 \\
        0.5 & 1
    \end{array}   \right]. \label{16}
\end{equation} Then, the R-D loss can be rewritten as
\begin{equation}
    \Delta J \approx \frac{1}{2} E\left[ \Delta \boldsymbol{x}^2 +\Delta \boldsymbol{x} \Delta \boldsymbol{w} + \Delta \boldsymbol{w}^2 \right]. \label{17}
\end{equation} If we happen to have a case with a larger quantization error, e.g., \ $[\Delta \boldsymbol{x},\Delta \boldsymbol{w}]$ \  = \  $[0.4,\ -0.4]$, and the other scenario with a smaller quantization error  \  $[0.3,\ 0.3]$,  the R-D loss is \ $0.08$\  and $0.135$, respectively. In this example, a larger quantization error gives a smaller R-D loss. 

{A similar observation was also discussed in AdaRound \cite{nagel2020up}, where their theoretical analysis only considered the weight quantization but ignored the impact of activation quantization.}

\begin{figure}[t]
    \centering
    \includegraphics[width=0.9\columnwidth]{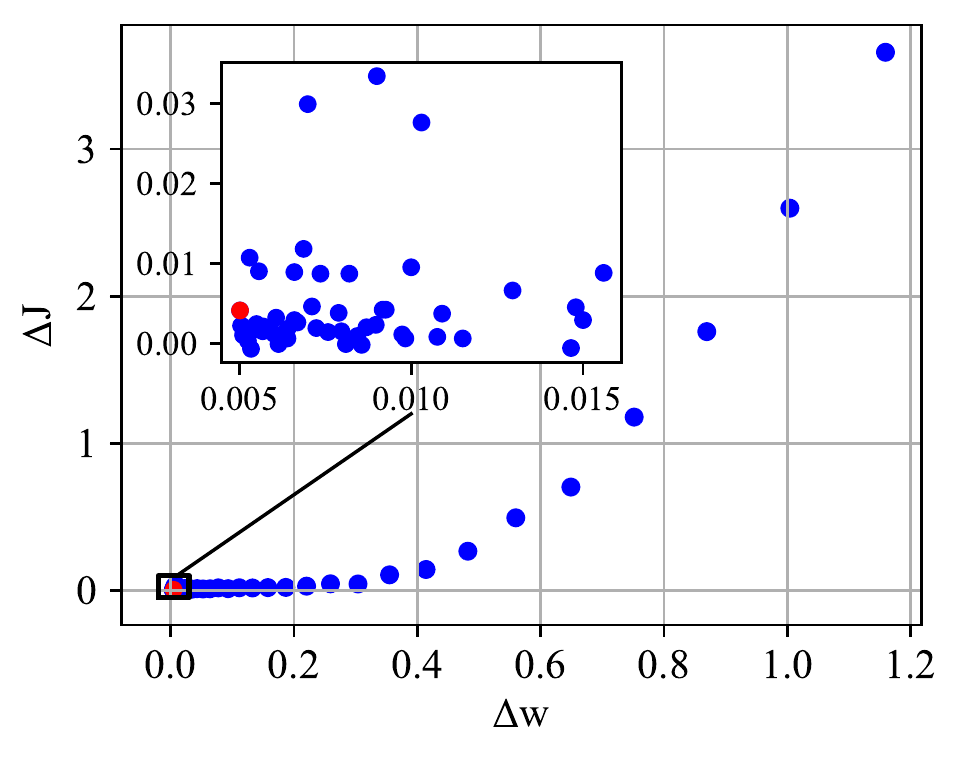}
    \caption{{\it Compression Task Loss vs. Quantization Error.} Localized non-monotonic behavior is presented where the minimization of quantization error $\Delta \boldsymbol{w}$ does not lead to the minimal loss of  rate-distortion metric ($\Delta J$).}
    \label{fig2}
\end{figure}

{\bf Localized Non-monotonic Behavior.} In the meantime, we visualize the absolute quantization error of weight, e.g., $\Delta \boldsymbol{w}$ with compression task loss measured by the $\Delta J$, further confirming the theoretical justification above. In Fig.~\ref{fig2}, although over a wide range of $\Delta \boldsymbol{w}$, it is monotonically related to the R-D loss $\Delta J$; it presents localized non-monotonic behavior where the minimization of quantization error does not lead to the minimum of $\Delta J$, which, consequently, degrades the overall compression performance of the underlying floating-point LIC model if we pursue the minimization of quantization error for model quantization.

Assuming the fixed activation, given a single element $\Delta w_i$ in $\Delta \boldsymbol{w}\in \mathbb{R}^{O\times I\times H\times W}$ with $O$/$I$ as the channel number of output/input tensor and $W$/$H$ as the width/height of convolutional kernel,  when $\Delta w_i$ is small, e.g., $\Delta w_i < 0.1$, $\Delta J$ is determined by $\Delta w_i^2$ and  $\Delta w_i$ jointly as in Eq. \eqref{eq:expanded_simp_task_loss}. In practice,  $\Delta w_i$ can take negative, zero, or positive values, making the sum of $\frac{\partial ^2 \boldsymbol{J}}{\partial x \partial w_i} \Delta x \Delta w_i$ +  $\frac{\partial ^2 \boldsymbol{J}}{\partial w_i^2} \Delta w_i^2$ fluctuate non-monotonically, as enlarged in Fig. \ref{fig2}. On the other hand, with the magnitude increase of quantization error $\Delta w$, the second-order term $\Delta w^2$ dominates, yielding the global monotonic behavior as also pictured in Fig. \ref{fig2}. Similar explanations can be made for activation parameters as well.

As also reported in Fig.~\ref{fig:quantize w} and \ref{fig:quantize w&a} experimentally, simply using MSE measures in model quantization by ignoring such localized non-monotonic behavior will lead to significant compression performance degradation when compared with the original floating-point model.

\subsection{Rate-Distortion Optimization}
As for the compression task, the optimization target in PTQ is to minimize the R-D loss by considering the distortion $D$ and bite rate $R$ jointly~\cite{Rate-distortion-Theory}. For a typical VAE structure with hyperprior, its R-D loss is 
\begin{align}
    J &= \lambda \cdot D + R_{\hat{\boldsymbol{Y}}} + R_{\hat{\boldsymbol{Z}}} \label{eq:R-D-loss0} \\
    &= \lambda \cdot D(\boldsymbol{X}, \hat{\boldsymbol{X}}) \nonumber\\
       &+ E[-\log_2(p_{\hat{\boldsymbol{Y}}}(\hat{\boldsymbol{Y}}|\hat{\boldsymbol{Z}}))] 
       + E[-\log_2(p_{\hat{\boldsymbol{Z}}}(\hat{\boldsymbol{Z}}))], \label{eq:R-D-loss}
\end{align} 
where $R = R_{\hat{\boldsymbol{Y}}} + R_{\hat{\boldsymbol{Z}}}$ is the total bit rate consumed by the latent feature $\hat{\boldsymbol{Y}}$ and side information $\hat{\boldsymbol{Z}}$, and $D$ is the distortion between the original $\boldsymbol{X}$ and its reconstruction $\hat{\boldsymbol{X}}$, which is typically computed using MSE or MS-SSIM. Besides, $p_{\hat{\boldsymbol{Y}}}$ and $p_{\hat{\boldsymbol{Z}}}$ are the probability distributions of respective $\hat{\boldsymbol{Y}}$ and $\hat{\boldsymbol{Z}}$ for entropy coding. We adapt Lagrange multiplier $\lambda$ for various bit rates as they do in~\cite{minnenbt18,cheng2020learned,TIC}.

For a typical VAE structure with hyperprior, we have:
\begin{align}
    \hat{\boldsymbol{Y}} = Round(\boldsymbol{Y}) = Round(g_a(\boldsymbol{X})), \label{eq:ga} \\
    \hat{\boldsymbol{Z}} = Round(\boldsymbol{Z}) = Round(h_a(\boldsymbol{Y})), \label{eq:ha}
\end{align}
where $g_a$ and $h_a$ are the main encoder and hyper encoder. Then latent feature $\hat{\boldsymbol{Y}} $ and side information $\hat{\boldsymbol{Z}}$ are used to reconstruct better $\hat{\boldsymbol{X}}$ at the minimum bitrate cost with the help of the main decoder $g_s$ and hyper decoder $h_s$. Thus, $\hat{\boldsymbol{X}}$ is conditioned on $\hat{\boldsymbol{Y}} $ and $\hat{\boldsymbol{Z}}$.  As a result, both rate term $R$ and distortion term $D$ are related to latent feature $\hat{\boldsymbol{Y}}$ and side information $\hat{\boldsymbol{Z}}$.

A basic demand in LIC is to achieve higher reconstruction quality at lower bit rates. Thus, we must balance the distortion term $D$ and rate term $R$ in quantization for a given floating-point model by optimizing \eqref{eq:R-D-loss}. The model's overall performance is compromised if we only consider distortion or rate separately. Therefore, joint Rate-Distortion Optimization is necessary.

{\bf Scaling Optimization.}
For a pretrained floating-point LIC model with R-D loss $J_0$, the examination of the proposed RDO-PTQ is defined below to derive the proper scaling factor for the quantization of activation, weight, and bias, e.g.,
\begin{align}
    s_{\boldsymbol{x}}, &s_{\boldsymbol{w}} ,s_{\boldsymbol{b}} \nonumber\\
    &= {\mathop{\arg \min}_{ s_{\boldsymbol{x}}, s_{\boldsymbol{w}} ,s_{\boldsymbol{b}}}} \ \lVert \widehat{J} - J_0 \rVert^2 \nonumber\\
             &= {\mathop{\arg \min}_{ s_{\boldsymbol{x}}, s_{\boldsymbol{w}} ,s_{\boldsymbol{b}}}} \ \lVert (\widehat{R} + \lambda \cdot \widehat{D}) - (R_0 + \lambda \cdot D_0) \rVert^2, \label{eq:RDO-PTQ}
\end{align}
where $\widehat{J}$ is the R-D measure of its fixed-point counterpart. The above loss function represents the global model task loss. We call it $\mathcal{L}_{task}$, i.e.,
\begin{equation}
    \mathcal{L}_{task} = \lVert \widehat{J} - J_0 \rVert^2 = \lVert (\widehat{R} + \lambda \cdot \widehat{D}) - (R_0 + \lambda \cdot D_0)\rVert^2. \label{eq:task_loss}
\end{equation}

However, optimizing all LIC model layers at once is impractical because the accumulation of quantization errors from layer to layer {makes it extremely difficult to quickly and reliably optimize such a great amount of parameters (e.g., weight, bias, and activation).} On the other hand, we only have a small calibration dataset (e.g., ten images in quantizing LIC) in PTQ, which may lead to over-fitting if all parameters are optimized together. As \cite{jakubovitz2019generalization} explained, the networks can have perfect expressivity when the number of parameters exceeds the number of data samples during training, but lower training error does not assure lower test error. 

We, therefore, suggest applying the layer-wise (e.g., per convolution layer) quantization to optimize the parameters progressively. Parameters at the former layer are fixed to optimize successive layers. As for the parameter quantization at the $l$-th layer, all parameters from the very first to the $(l-1)$-th layer are already quantized and fixed for optimization, while floating-point parameters from the $(l+1)$-th layer to the last layer are kept unchanged. 
Thus, Eq.~(\ref{eq:RDO-PTQ}) is reformulated as:
\begin{equation}
    s^l_{\boldsymbol{x}}, s^l_{\boldsymbol{w}} ,s^l_{\boldsymbol{b}} 
    = {\mathop{\arg \min}_{s^l_{\boldsymbol{x}},s^l_{\boldsymbol{w}},s^l_{\boldsymbol{b}}}} \ \lVert \widehat{J}(\hat{{\boldsymbol{x}}}^l, \hat{\boldsymbol{w}}^l, \hat{\boldsymbol{b}}^l) - J_0(\boldsymbol{x}^l,\boldsymbol{w}^l,\boldsymbol{b}^l) \rVert^2, \label{eq:scale_layer}
    \end{equation} where
$\hat{\boldsymbol{x}}^l = \Lambda( \hat{\boldsymbol{w}}^{l-1}\hat{\boldsymbol{x}}^{l-1}+ \hat{\boldsymbol{b}}^{l-1})$  with $\Lambda(\cdot)$ as a nonlinear activation function like ReLU. 

{\bf Rounding Optimization.}
Recalling the quantization in Eq.~\eqref{1}, {it relates to scaling factors and the rounding error.} Usually, rounding-to-nearest (i.e., $\lfloor \cdot \rceil$) is the most widely used. However,  recent works (e.g., AdaRound \cite{nagel2020up}, AdaQuant \cite{hubara2020improving}, BRECQ \cite{li2020brecq}) have shown that rounding-to-nearest is not optimal and then suggest the adaptive rounding strategies. Specifically, weight $\boldsymbol{{w}}$ is initially rounded to $\boldsymbol{\widehat{w}}$ through the use of floor rounding (i.e., $\lfloor \cdot \rfloor$), and a learnable variable $\boldsymbol{V}$ is then added to quantized weight $\boldsymbol{\widehat{w}}$ for adaptive rounding. 

Accordingly, the optimization objective is written as:
\begin{equation}
    {\mathop{\arg \min}_{\boldsymbol{V}}} \ \lVert \Lambda( \boldsymbol{w^lx^l}) - \Lambda( \boldsymbol{\widehat{w}^lx^l}) \rVert^2 + \lambda f_{reg}(\boldsymbol{V}), \label{eq:rec+reg}
    \end{equation}
where $\Lambda(\cdot)$ is the activation function and $\lambda f_{reg}(\boldsymbol{V})$ is a differentiable regularization term introduced to encourage $\boldsymbol{\widehat{w}}$ convergence. Note that \eqref{eq:rec+reg} was originally used in AdaRound.

We then extend per-layer quantization loss $\mathcal{L}_{lq}$ in Eq.~\eqref{eq:rec+reg} to take both quantized activation $\boldsymbol{\widehat{x}}$ and weight $\boldsymbol{\widehat{w}}$ into consideration jointly, i.e.,
\begin{equation}
    \mathcal{L}_{lq} = \lVert \Lambda( \boldsymbol{w^lx^l}) - \Lambda( \boldsymbol{\widehat{w}^l \widehat{x}^l})\rVert^2 + \lambda f_{reg}(\boldsymbol{V}). \label{eq:rec_loss}
\end{equation}  Note that the same rules used for the weight can also be applied to the bias.

In the end, Eq.~\eqref{1} is reformatted by jointly optimizing the scaling and rounding. Thus, in the proposed RDO-PTQ, we have:
\begin{equation}
    \boldsymbol{s}, \boldsymbol{V} = \mathop{\arg \min}_{\boldsymbol{s}, \boldsymbol{V}} \lambda_t \mathcal{L}_{task} + \mathcal{L}_{lq},
\end{equation}
where $\lambda_t$ is a hyper-parameter to make a trade-off between two losses. We observe that experimental results are not sensitive to $\lambda_t$. For simplicity, $\lambda_t$ is set to $1$.

In this way, the optimization will jointly consider per-layer quantization loss and task loss. 

\subsection{Dynamic Range Determination} 
In Eq.~(\ref{3}), when the bit width precision is given, the scaling factor, e.g., $s_{\boldsymbol{w}}^l$, $s_{\boldsymbol{b}}^l$, or $s_{\boldsymbol{x}}^l$, is only related to the dynamic range $r^l$ of the corresponding data tensor, e.g., ${\boldsymbol{w}}^l$, ${\boldsymbol{b}}^l$, or ${\boldsymbol{x}}^l$. For the uniform quantization used mostly, the derivation of the scaling factor is equivalent to the determination of the dynamic range. In general, weight, bias, and activation assume a similar Gaussian distribution~\cite{hong2020efficient}, and their ranges are approximated below. 

\begin{figure}[t]
\centering
\subfigure[weight distribution]{
\includegraphics[width=0.46\columnwidth]{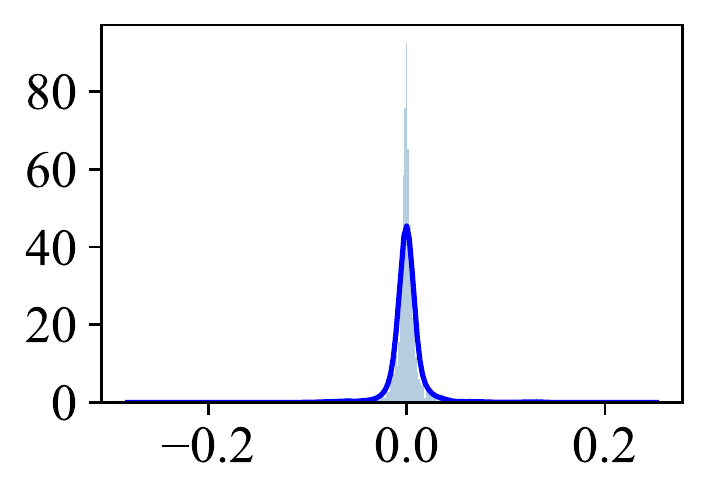} 
\includegraphics[width=0.48\columnwidth]{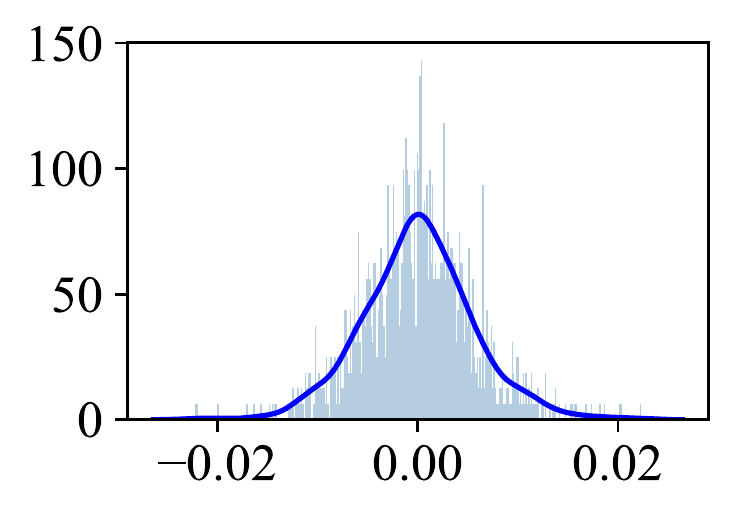} 
}
\\

\subfigure[activation distribution]{
\includegraphics[width=0.45\columnwidth]{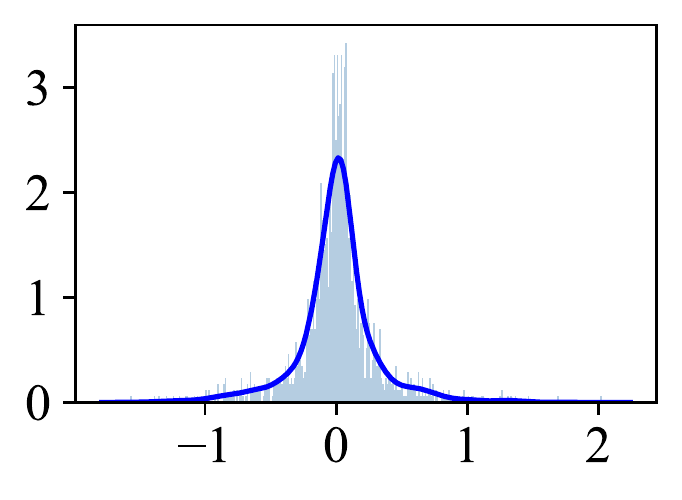} 
\includegraphics[width=0.47\columnwidth]{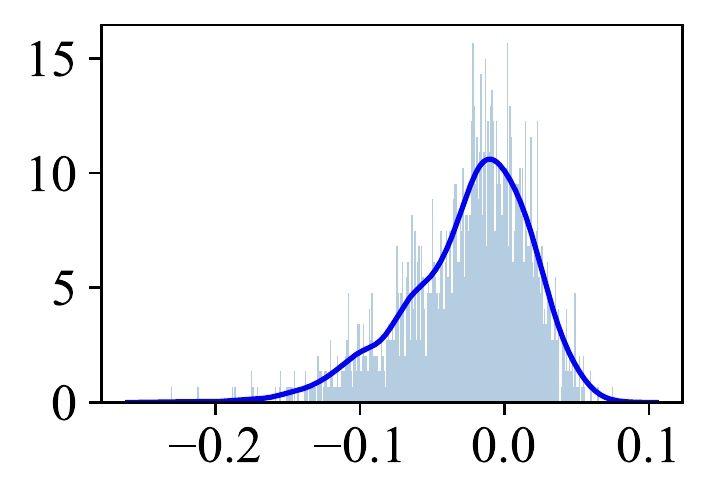} 
}

\caption{Exemplified weight and activation distribution of two channels for a given layer. Other layers exhibit similar distributions from one channel to another. The horizontal axis represents the intensity level range of the underlying parameter (e.g., weight or activation), and the vertical axis shows the probability at different intensity levels of the parameter.  The blue solid line approximates the probability density function of the parameter.}
\label{eq:channel_wise_weight}
\end{figure}

{\bf Weight.} As shown in Fig.~\ref{eq:channel_wise_weight}, weight distribution varies from one layer to another of a given LIC model and from one channel to another at a given layer. This suggests modeling the range of the weight tensor using
\begin{align}
    r_{\boldsymbol{w}}^{l,k} = N_{\boldsymbol{w}}^{l,k}\cdot (\max( {\boldsymbol{w}}^{l,k} ) - \min( {\boldsymbol{w}}^{l,k})), \label{eq:weight_range}
\end{align} with $\max( {\boldsymbol{w}}^{l,k} ) - \min( {\boldsymbol{w}}^{l,k})$ directly computed from $k$-th channel weights at $l$-th layer. As seen, {our method properly determines per-channel $N_{\boldsymbol{w}}^{l,k}$ for weight quantization.} Although Sun et al.~\cite{sun2021learned} did a similar channel-wise grouping to clip weights for fixed-point processing, model parameters retraining was required for finetuning, which largely differs from our PTQ solution. 

{\bf Activation.} Feature activations are closely related to the original image input. However, Hong et al.~\cite{hong2020efficient} enforced a fixed dynamic range per layer, e.g., [$\mu - 3\sigma$, $\mu + 3\sigma$] to normalize the activation, which is apparently sub-optimal without thoroughly considering the dynamics of input content. Hong et al.~\cite{hong2022daq} adopted a distribution-aware parameter $\gamma$, e.g., [$\mu - \gamma\sigma$, $\mu + \gamma\sigma$] as the quantization range, but channel-wisely calculating the mean $\mu$ and variance $\sigma$ is complicated.

After carefully examining the activation distribution, the channel-wise variations in the activation tensor are also considered, leading to 
\begin{align}
    r_{\boldsymbol{x}}^{l,k} = N_{\boldsymbol{x}}^{l,k} \cdot (\max({\boldsymbol{x}}^{l,k})-\min({\boldsymbol{x}}^{l,k})). \label{eq:activation_range}
\end{align}

{\bf Bias.} First, the range of the bias is approximated layer-wisely using
\begin{align}
        r_{\boldsymbol{b}}^{l} = N_{\boldsymbol{b}}^{l}\cdot (\max({\boldsymbol{b}}^{l})-\min({\boldsymbol{b}}^{l}) ), \label{eq:bias_range}
\end{align} to determine the scaling factor $s_{\boldsymbol{b}}^l$ using Eq. (\ref{3}).

Typically, the bias term is augmented with the product of weight and activation to output feature activation of the current $l$-th layer as the input of the next $(l+1)$-th layer. Even though we enforce INT8 precision for all data tensors, in practice, a 32-bit accumulator is often used to host intermediate data $\hat{\boldsymbol{x}}_{imd}^{l}$ to avoid potential data overflow. This suggests:
\begin{align}
    \hat{\boldsymbol{x}}_{imd}^{l} = \hat{\boldsymbol{w}}^l\cdot\hat{\boldsymbol{x}}^l + \hat{\boldsymbol{b}}^l.
\end{align} $\hat{\boldsymbol{x}}_{imd}^{l}$ needs to be re-quantized from 32-bit to 8-bit so that it can be passed to the next layer without overflow. Generally, this operation is complex. Here, we wish to simply scale 32-bit $\hat{\boldsymbol{x}}_{imd}^{l}$ to have 8-bit $\hat{\boldsymbol{x}}^{l+1}$ to input the $(l+1)$-th layer.

By combining Eq. (\ref{2}), as long as  we have 
\begin{align}
    \tilde{\boldsymbol{b}}_{int}^{l} = \lfloor \frac{s_{\boldsymbol{b}}^l}{s_{\boldsymbol{w}}^l \cdot s_{\boldsymbol{x}}^l} \cdot {\boldsymbol{b}_{int}^l} \rceil.  \label{eq:bias_rescaling}
    \end{align} we then arrive at 
\begin{align}
     \hat{\boldsymbol{x}}_{imd}^{l}  &= s_{\boldsymbol{w}}^l\cdot{\boldsymbol{w}}_{int}^l\cdot s_{\boldsymbol{x}}^l\cdot{\boldsymbol{x}}_{int}^l + s_{\boldsymbol{b}}^l\cdot{\boldsymbol{b}}_{int}^l \nonumber\\
     & \approx s_{\boldsymbol{w}}^l\cdot s_{\boldsymbol{x}}^l\cdot({\boldsymbol{w}}_{int}^l\cdot{\boldsymbol{x}}_{int}^l+ \tilde{\boldsymbol{b}}_{int}^l), \label{eq:activation_transfer}
\end{align} which shows that  we can simply scale 32-bit $\hat{\boldsymbol{x}}_{imd}^{l}$ to derive 8-bit input ${\boldsymbol{x}}_{int}^{l+1}$, i.e.,
        ${\boldsymbol{x}}_{int}^{l+1} = \frac{\hat{\boldsymbol{x}}_{imd}^{l}}{s_{\boldsymbol{x}}^{l+1} }$. 
        
Often time,  $\frac{s_{\boldsymbol{b}}^l}{s_{\boldsymbol{w}}^l \cdot s_{\boldsymbol{x}}^l}\gg 1$ in Eq. (\ref{eq:bias_rescaling}) which would not affect the task performance due to rounding operation. We call the operation defined in Eq. (\ref{eq:bias_rescaling}) bias rescaling, which is then used to enforce strict INT8 computation layer by layer.

On the contrary, many existing works either assume the INT32 precision or brutally enforce zeros for bias. In a neural network, bias controls the neuron activation state.  Inappropriate processing of bias may lead to catastrophic results. For example, having the bias in INT32 precision may cause data overflow of the underlying accumulator, while setting the bias to zero would degrade the model performance significantly. {For example, simply discarding the bias would lead to an average PSNR drop of more than 12 dB for quantized Lu2022 on the Kodak dataset~\cite{TIC} (when compared with its floating-point model).} 

\subsection{Summary}

The overall RDO-PTQ process is summarized in Algorithm \ref{algorithm}, which is a calibration process to determine proper quantization parameters (e.g., scaling factors) for quantizing model parameters (e.g., weight, bias) and activation.  These quantization parameters are updated by backward propagation and gradient descent in optimization, where we use R-D loss but not the rate or distortion only as a target to optimize them. More details can be found in the supplementary material.

We optimize the model from one network layer to another. When the loss $\mathcal{L}$ converges, or the iteration ends, the optimization for the $l$-th layer is completed, and all settings are fixed for the optimization of the $(l+1)$-th layer. Such a process continues until the last layer of the model. 
Note that except for the bias that we apply the layer-wise processing with rescaling, we perform channel-wise quantization for both weight and activation parameters. 

In training, we approximate the rounding operation in (\ref{eq:ga}) and (\ref{eq:ha}) by adding uniform noise to ensure the differentiability in backpropagation, as in \cite{ballemshj18, balle2018integer, minnenbt18}.



\begin{algorithm}[tb]
    \caption{RDO-PTQ for Learned Image Compression}
    \label{algorithm}
    \renewcommand{\algorithmicrequire}{\textbf{Input:}} 
    \renewcommand{\algorithmicensure}{\textbf{Output:}}
    \begin{algorithmic}[1]
        \REQUIRE Floating-point model; Calibration image set\\
        
        \FOR{$l$ in all layers $\{ L \}_{l=1}^N$  of FP model}
            \STATE Initial quantization parameters of $\boldsymbol{x}^l$, $\boldsymbol{w}^l$, $\boldsymbol{b}^l$;
            
            \REPEAT
                \STATE Quantize $l$-th layer;
                \STATE Forward propagation;
                \STATE Compute $\mathcal{L}_{task}$, $\mathcal{L}_{lq}$;
                \STATE Backward propagation;
                \STATE Update quantization parameters and rounding of weights by gradient descent\cite{hinton2012neural, kingma2014adam};
            \UNTIL{Convergence or excess limitation}
        \ENDFOR
        \ENSURE The quantized model; Optimized quantization parameters
    \end{algorithmic}

\end{algorithm}

\section{Experiments} \label{sec:exp}

In this section, extensive experiments are conducted to report the efficiency of the proposed RDO-PTQ for LIC model quantization. We first describe the experimental settings, then the evaluations and analyses of various models with different methods are presented in detail.

\subsection{Comparison Setup}

\begin{figure*}[htb]
\vspace{-0.4cm}
\centering
\subfigure[]{
\includegraphics[width=0.43\textwidth]{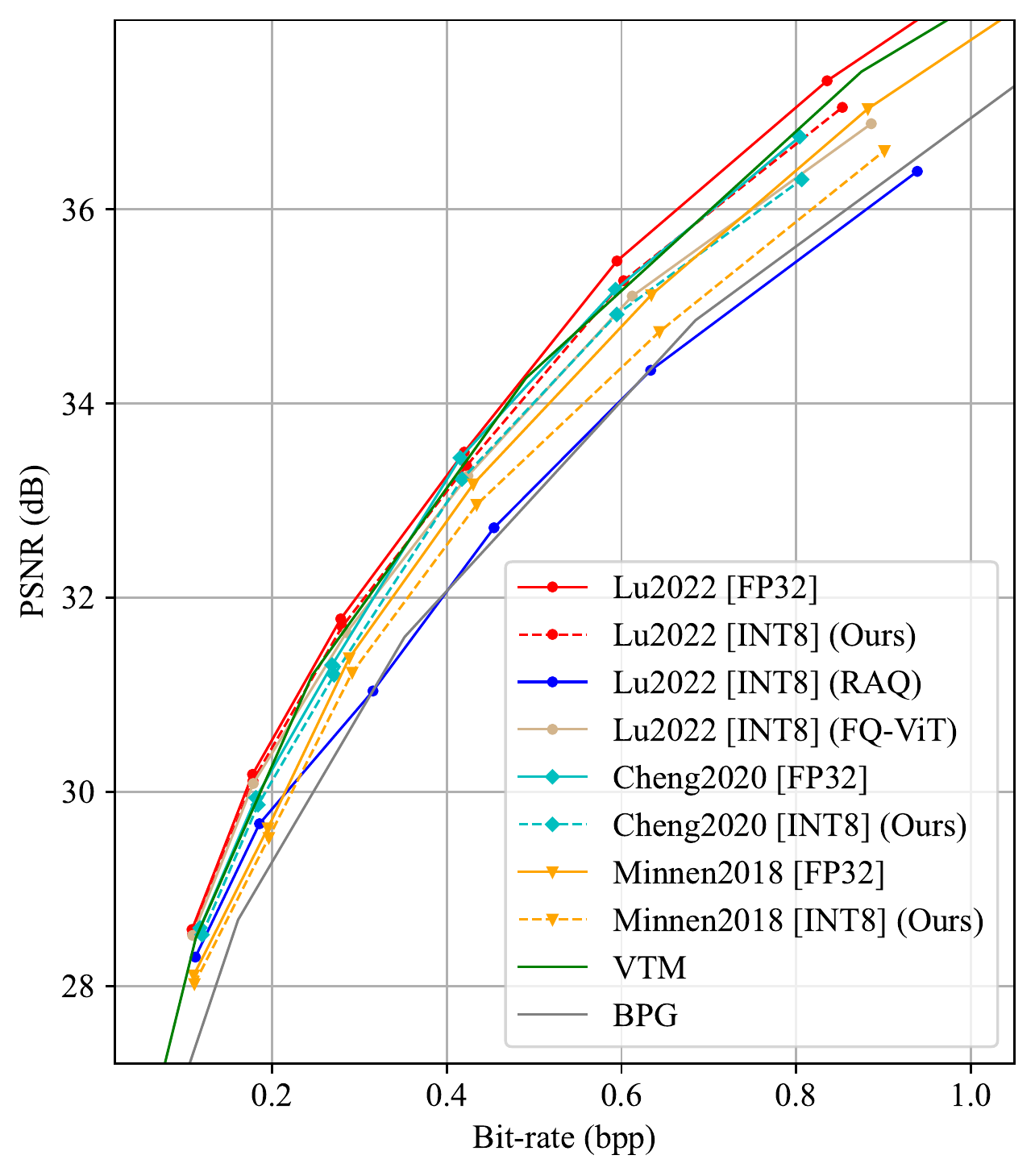} 
}
\hspace{0.8cm}
\subfigure[]{
\includegraphics[width=0.43\textwidth]{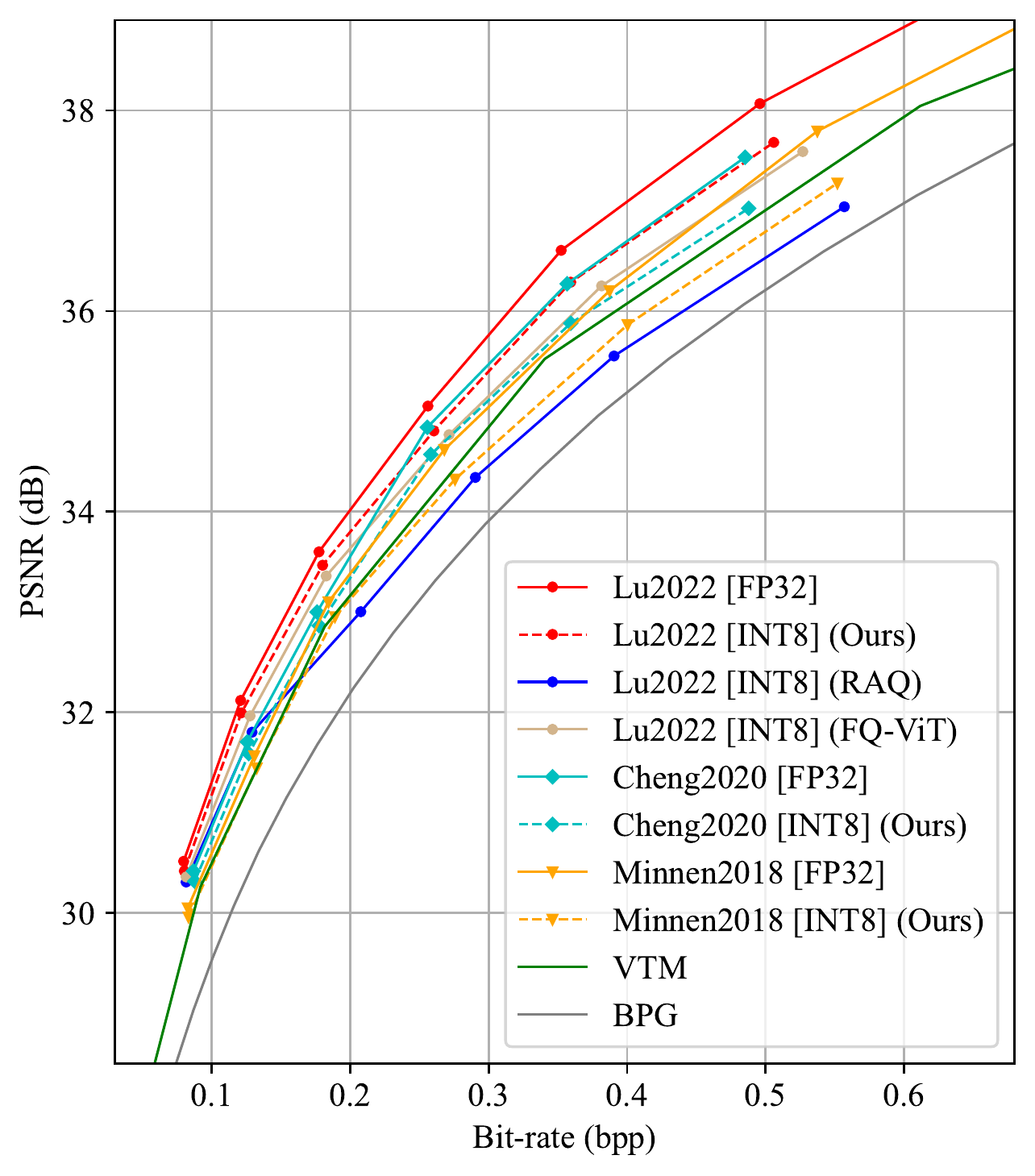} 
}
\caption{{\bf R-D Performance.} (a) Kodak dataset; (b) Tecnick dataset; VTM stands for the VVC Intra, while BPG is the HEVC Intra. Methods marked with [FP32] represent their original 32-bit floating-point models, while methods highlighted with [INT8] are quantized models using 8-bit fixed-point precision for processing.}
\label{RD curves}
\end{figure*}

{\bf Pretrained FP32 LICs.} We choose three popular LIC models in their native FP32 format for evaluation, namely Minnen2018~\cite{minnenbt18}, Cheng2020~\cite{cheng2020learned}, and Lu2022~\cite{TIC}.
For instance, Minnen2018 is a seminal work first introducing the joint exploration of hyperprior and autoregressive neighbors for entropy coding of latent features, and Cheng2020 is one of several works that first proposed to apply attention mechanism. Both Minnen2018 and Cheng2020 rely on stacked convolutions, while Lu2022 is one of several earlier attempts that uses convolution and local attention Vision Transformer~\cite{hassani2023neighborhood}. Subsequent results report the generalization of our RDO-PTQ to these representative models regardless of its key unit (e.g., convolution or self-attention, ReLU or GDN, etc.), which is attractive to practical applications. 

Pretrained models are directly obtained from their open-source websites. The models used in the comparison are trained using MSE and MS-SSIM loss for distortion measurement in the R-D metric. For each model, six bitrates are experimented with by setting six different $\lambda$s. As for MSE loss used for distortion measurement, $ \lambda$ is chosen from $\{0.0018,\ 0.0035,\ 0.0067,\ 0.013,\ 0.025,\ 0.0483\}$; while for MS-SSIM loss used for distortion measurement, $\lambda$ is chosen from $\{2.40,\ 4.58,\ 8.73,\ 16.64,\ 31.73,\ 60.50\}$. Here, the actual loss function in Eq. \eqref{eq:R-D-loss0} is $\lambda \cdot 255^2 \cdot D + R$ for MSE loss trained models and $\lambda \cdot (1-D) + R$ for MS-SSIM loss trained models.

{\bf Parametric PTQs.} Given that Lu2022 currently demonstrates the leading compression efficiency, we mainly use it as the baseline model to implement other model quantization schemes.  Unfortunately, to the best of our knowledge, we are unaware of any open-source PTQ method specifically for quantizing LIC models. For a fair comparison, we implement the Range-Adaptive Quantization (RAQ) \cite{hong2020efficient} initially requiring model retraining as a PTQ approach. On the other hand, we also include the FQ-ViT \cite{lin2022fqvit} for comparative study. It is a PTQ method initially designed for image classification and objective detection using the Transformer backbone. Here, we extend its main idea to support the compression task. RAQ and FQ-ViT determine the scaling factors according to the parameter distribution, which does not assure optimality regardless of the MSE or task loss.

{\bf MSE-optimized PTQ} is studied to understand the potential of R-D loss used in LIC model quantization, which is defined by Eq.~\eqref{eq:MMSE}.

{\bf Task-oriented PTQs.}  We further make comparisons with AdaRound \cite{nagel2020up} and BRECQ \cite{li2020brecq}, which are two PTQ methods developed recently. Different from our method that measures the compression task (or classification) using the R-D (accuracy) metric explicitly, both AdaRound \cite{nagel2020up} and BRECQ \cite{li2020brecq} suggest the use of second-order approximation to model the task loss implicitly. Note that only layer-wise activation quantization is considered in AdaRound and BRECQ. For a fair comparison, we enforce the same channel-wise activation quantization for low-level tasks while the same layer-wise activation quantization for high-level tasks. We also include the Q-LIC~\cite{sun2022q} in this category.

{\bf QATs.} 
Another two QAT methods developed in \cite{sun2020end, sun2021learned} that require model retraining for LIC quantization are also evaluated for comparative study.

{\bf Testing Datasets.} Three popular datasets that contain diverse images are used for evaluation, i.e., the Kodak dataset with an image size of $768\times512$, the Tecnick dataset with an image size of $1200\times1200$, and the CLIC professional validation dataset, which contains 41 images with 2k spatial resolution approximately. The Peak Signal-to-Noise Ratio (PSNR) and MS-SSIM are used to measure the image quality, and the bits per pixel (bpp) report the consumption of compressed bitrate.

\subsection{Evaluation}

{\bf Quantitative Performance.} We plot R-D curves in Fig. \ref{RD curves} for various floating-point LIC models and their quantized INT8 counterparts, and further report the BD-rate performance \cite{bjontegaard2001calculation} over 32-bit floating-point models for different PTQ approaches in Table \ref{tab:BD-rates}. In the meantime, compression efficiency using VVC Intra (VTM) and HEVC Intra (BPG) is also exemplified to help understand the relative performance gaps between quantized LICs and traditional image coders.

\begin{table}[htbp]
  \centering
  \caption{BD-rate loss over floating-point models. We highlight the best results in each block in \pmb{bold}.}
    \begin{tabular}{clrr}
    \toprule
    \multirow{2}[4]{*}{Model} & \multicolumn{1}{c}{\multirow{2}[4]{*}{Method}} & \multicolumn{2}{c}{Datasets} \\
\cmidrule{3-4}          &       & \multicolumn{1}{l}{Kodak} & \multicolumn{1}{l}{Tecnick} \\
    \midrule
    \multirow{3}[6]{*}{Lu2022 \cite{TinyLIC}} & FQ-ViT \cite{lin2022fqvit} & 7.06\% & 12.13\% \\
\cmidrule{2-4}          & RAQ \cite{hong2020efficient}   & 29.40\% & 31.04\% \\
\cmidrule{2-4}          & Ours   & \pmb{3.70\%} & \pmb{6.21\%} \\
    \midrule
    \multirow{2}[4]{*}{Cheng2020 \cite{cheng2020learned}} & RAQ \cite{hong2020efficient}   & 27.84\% & 29.95\% \\
\cmidrule{2-4}          & Ours  & \pmb{4.88\%} & \pmb{6.86\%} \\
    \midrule
    \multirow{2}[4]{*}{Minnen2018 \cite{minnenbt18}} & RAQ \cite{hong2020efficient}   & 30.41\% & 31.55\% \\
\cmidrule{2-4}          & Ours  & \pmb{5.84\%} & \pmb{8.23\%} \\
    \bottomrule
    \end{tabular}%
  \label{tab:BD-rates}
\end{table}%

As seen, quantized INT8 LICs using our RDO-PTQ provide the least BD-rate loss to corresponding floating-point LIC models, greatly outperforming other PTQ alternatives (see averaged results in Table~\ref{tab:BD-rates} and curves in Fig. \ref{RD curves}). For RAQ \cite{hong2020efficient}, the author applies layer-wise quantization without considering the layer's diverse channel distributions, which degrades the performance significantly. Table \ref{tab:BD-rates} shows that RDO-PTQ outperforms RAQ over $20\%$ BD-rate gains, validating the effectiveness of channel-wise quantization. More importantly, having the anchor of Lu2022 [FP32] and Lu2022 [INT8] using our RDO-PTQ provides similar performance as the VVC Intra on the Kodak dataset and largely outperforms it on the Tecnick dataset.

As visualized in Fig. \ref{RD curves}, the BD-rate loss enlarges at higher bitrates for the proposed RDO-PTQ on various floating-point LICs. One potential reason is the increase of channels for high-bitrate LIC models (e.g., from 192 to 320). Note that we optimize the PTQ channel-wisely, the accumulated quantization error is typically larger for the model with more channels. {Such a problem can be resolved by optimizing the PTQ for all layers simultaneously, which may require excessive computations and calibration datasets to stay generalizable.} This is an interesting topic for further study.

\begin{figure*}[t]
    \centering
    \subfigure[Weight only quantization.]{
        \includegraphics[width=0.85\columnwidth]{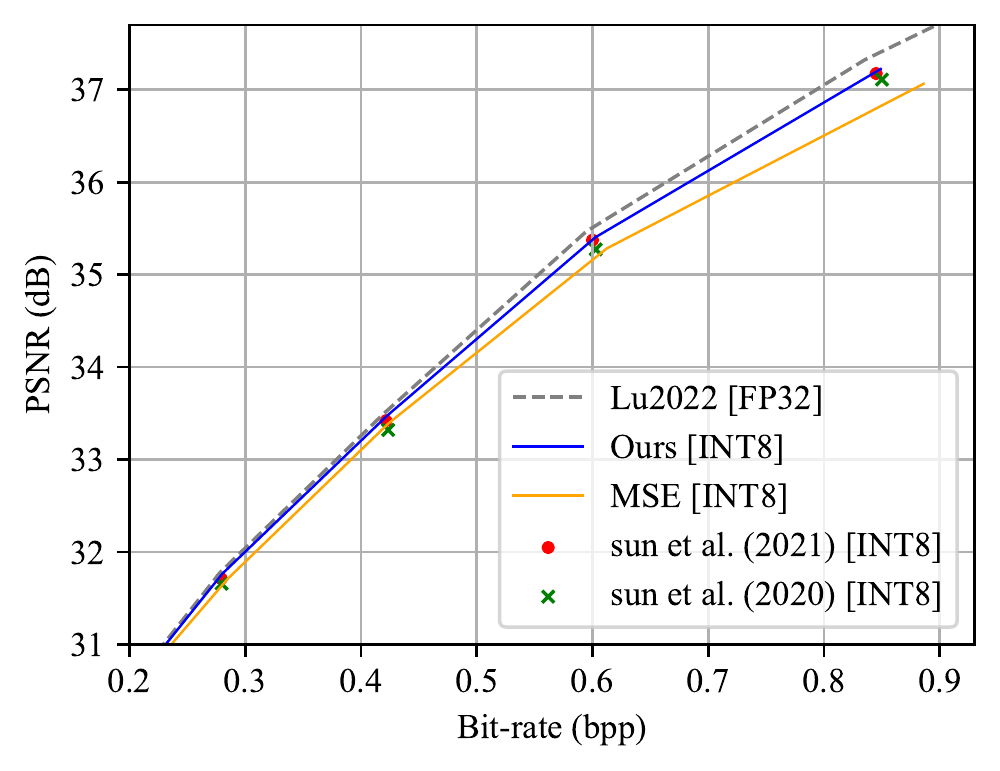} 
        \label{fig:quantize w}
    }
    \subfigure[Weight and activation quantization.]{
        \includegraphics[width=0.87\columnwidth]{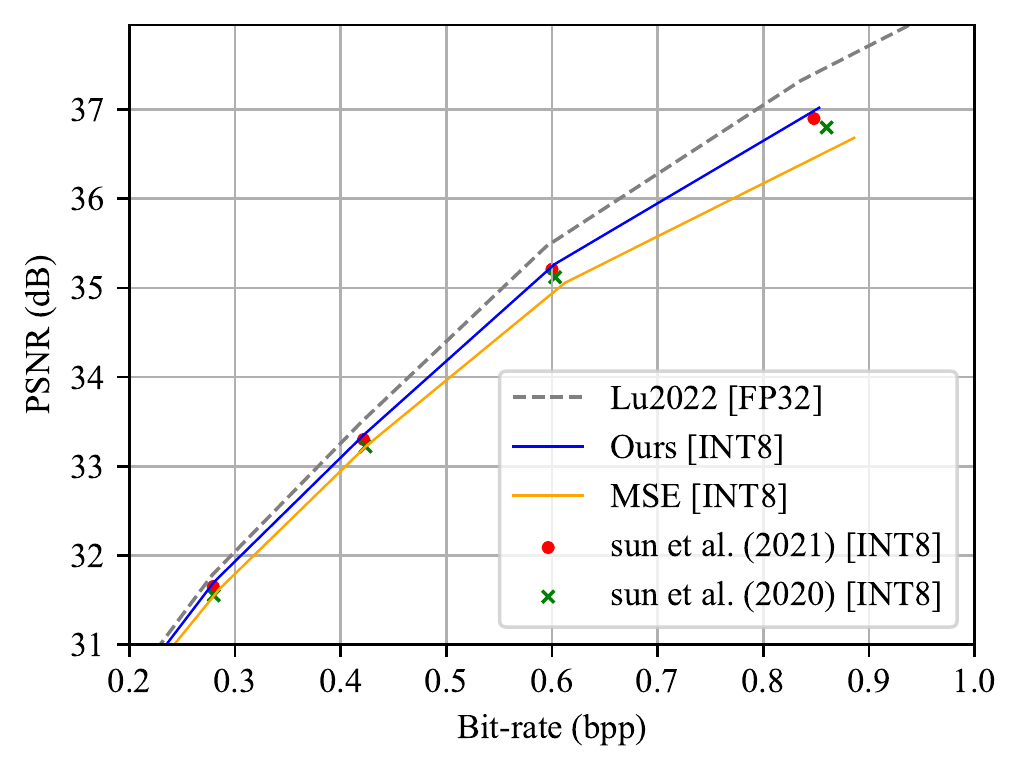}
        \label{fig:quantize w&a}
    }
    \caption{Comparison among different SOTA quantization strategies in LIC on Kodak.}
    \label{fig:qat_mse}
\end{figure*}

{\bf Comparison with Existing Methods.}
We first make a comparison with two QAT methods in Sun et al. (2020) \cite{sun2020end} and Sun et al. (2021) \cite{sun2021learned}. Given that there are no publicly accessible source codes, we faithfully reproduce them.
To retrain the model, the DIV2K dataset \cite{DIV2K} is used, and image samples are randomly cropped into fixed-size patches at $256 \times 256 \times 3$. 

In their work, both methods in \cite{sun2020end, sun2021learned} only quantize weights. 
Two experiments are conducted, i.e., weight-only quantization and weight and activation quantization. For the latter one,  we add the same channel-wise quantization of activations for a fair comparison.

Fig. \ref{fig:quantize w} shows the result of weight-only quantization, while Fig. \ref{fig:quantize w&a} shows the result for quantizing both weight and activation. The BD-rate losses of 8-bit quantized models over floating-point Lu2022 are $4.43\%$ and $6.78\%$ respectively, which are even higher than ours (e.g., 3.70\% and 6.21\% in Table~\ref{tab:BD-rates}), {not mentioning that our method does not require the model parameters retraining as these two methods~\cite{sun2020end,sun2021learned}, but instead only optimizing quantization parameters.}

As a comparative baseline,  MSE-optimized PTQ  is also plotted in Fig.~\ref{fig:qat_mse} as well, which provides the largest R-D loss as compared with the floating-point Lu2022, e.g., 7.82\% BD-Rate increase on average for 8-bit quantization of both weight and activation. When referring to the subplot in Fig.~\ref{fig:quantize w&a}, we notice that MSE-optimized PTQ works well at two lower bitrates (e.g., R-D points are mostly overlapped with those produced by Sun et al. (2020) and Sun et al. (2021)), but it presents a much larger loss at higher bitrates, leading to a noticeable gap to the proposed RDO-PTQ method.  Clearly, the smaller quantization errors do not lead to better compression performance. 

Sun et al.~\cite{sun2022q} introduced Q-LIC, a PTQ method to quantize the LIC model. They demonstrated that the influence of quantization error on the final reconstruction differed across channels. As a result, they manually categorized channels into different groups according to the quantization-incurred reconstruction errors.  Although they included the reconstruction error instead of the quantization error in PTQ optimization, they did not connect reconstruction performance with the weight, bias, and activation following the R-D optimization means, which is very different from our proposed RDO-PTQ. Unfortunately, we do not find any publicly accessible material on Q-LIC to perform the comprehensive comparison. Therefore, we use the data from its published paper for a simple comparison, to which we enforce our tests close to their settings.

The comparison between our method and Q-LIC \cite{sun2022q} is shown in Table \ref{tab:qlic}, where the floating-point model anchor is Cheng2020 \cite{cheng2020learned}. Our method largely outperforms Q-LIC \cite{sun2022q} in MSE loss trained models, while it also provides comparable performance to Q-LIC in MS-SSIM loss trained models. We notice that energy distribution is more concentrated on some specific channels for MS-SSIM loss trained models, which greatly helps Q-LIC to obtain lower reconstruction errors.
 

\begin{table}[t]
  \centering
  \caption{BD-rate loss over floating-point Cheng2020 on Kodak.}
    \begin{tabular}{lrr}
    \toprule
          & \multicolumn{1}{l}{MSE loss} & \multicolumn{1}{l}{MS-SSIM loss} \\
    \midrule
    Ours  & \pmb{4.88\%} & 4.65\% \\
    Q-LIC \cite{sun2022q}  & 10.50\% & \pmb{4.40\%} \\
    \bottomrule
    \end{tabular}%
  \label{tab:qlic}%
\end{table}%

Finally, we extend AdaRound \cite{nagel2020up} and BRECQ \cite{li2020brecq}, initially designed for high-level vision tasks, to support the quantization of the compression task. These methods assume the second-order approximation for task loss. Thus, they only use $\mathcal{L}_{lq}$ in optimization. And, neither AdaRound nor BRECQ considers the activation quantization during optimization. 
For a fair comparison, we implement the same channel-wise activation quantization. The results are listed in Table \ref{tab:ada+brecq}. Our method again reports the best performance, suggesting that explicitly optimizing the R-D metric in PTQ can better preserve the efficiency of the floating-point LIC model. Such an observation will also be validated for high-level tasks.

\begin{table}[t]
  \centering
  \caption{BD-rate loss over floating-point Lu2022. Various bit-width settings for the weight (w) and activation (a) are tested.}
    \begin{tabular}{cccccc}
    \toprule
    \multirow{2}[2]{*}{Method} & \multicolumn{2}{c}{bit-width} & \multirow{2}[2]{*}{Kodak} & \multirow{2}[2]{*}{Tecnick} & \multirow{2}[2]{*}{CLIC} \\
          & w     & a     &       &       &  \\
    \midrule
    AdaRound \cite{nagel2020up} & 6     & 32    & 10.08\% & 15.79\% & 13.25\% \\
    BRECQ \cite{li2020brecq} & 6     & 32    & 9.01\% & 14.14\% & 12.23\% \\
    Ours  & 6     & 32    & \pmb{7.92\%} & \pmb{12.02\%} & \pmb{11.28\%} \\
    \midrule
    AdaRound \cite{nagel2020up}  & 8     & 32    & 1.28\% & 2.58\% & 2.23\% \\
    BRECQ \cite{li2020brecq} & 8     & 32    & 1.18\% & 2.32\% & 2.01\% \\
    Ours  & 8     & 32    & \pmb{0.82\%} & \pmb{2.02\%} & \pmb{1.87\%} \\
    \midrule
    AdaRound \cite{nagel2020up}  & 8     & 8     & 4.96\% & 8.03\% & 7.94\% \\
    BRECQ \cite{li2020brecq} & 8     & 8     & 4.51\% & 7.75\% & 7.59\% \\
    Ours  & 8     & 8     & \pmb{3.70\%} & \pmb{6.21\%} & \pmb{6.16\%} \\
    \bottomrule
    \end{tabular}%
  \label{tab:ada+brecq}%
\end{table}%

{\bf Evaluation of 10-bit Quantization.} Previous discussions mainly focus on the 8-bit quantization. As visualized in plots, R-D loss enlarges as the bitrate increases. The same phenomena are observed in other works as well. This is attributed to the error accumulation from one layer to another in PTQ optimization.

Thus, adopting a higher bit-width is a potential solution, particularly for those application scenarios accepting mixed-precision implementation. We follow the studies by Hong et al. in~\cite{hong2020efficient} to examine the 10-bit quantization.


\begin{table}[bp]
  \centering
  \caption{BD-rate loss over various floating-point models under 10-bit quantization.}
    \begin{tabular}{ c c c c }
    \toprule
          & Lu2022 & Cheng2020 & Minnen2018 \\
    \midrule
    Kodak & 0.49\%  & 0.43\%  & 0.41\%  \\
    \midrule
    Tecnick & 1.03\%  & 0.65\%  & 1.22\%  \\
    \bottomrule
    \end{tabular}%
  \label{tab_appendix:bd-rate}%
\end{table}%

\begin{figure}[t]
    \centering
    \subfigure[Kodak dataset]{
        \includegraphics[width=0.43\textwidth]{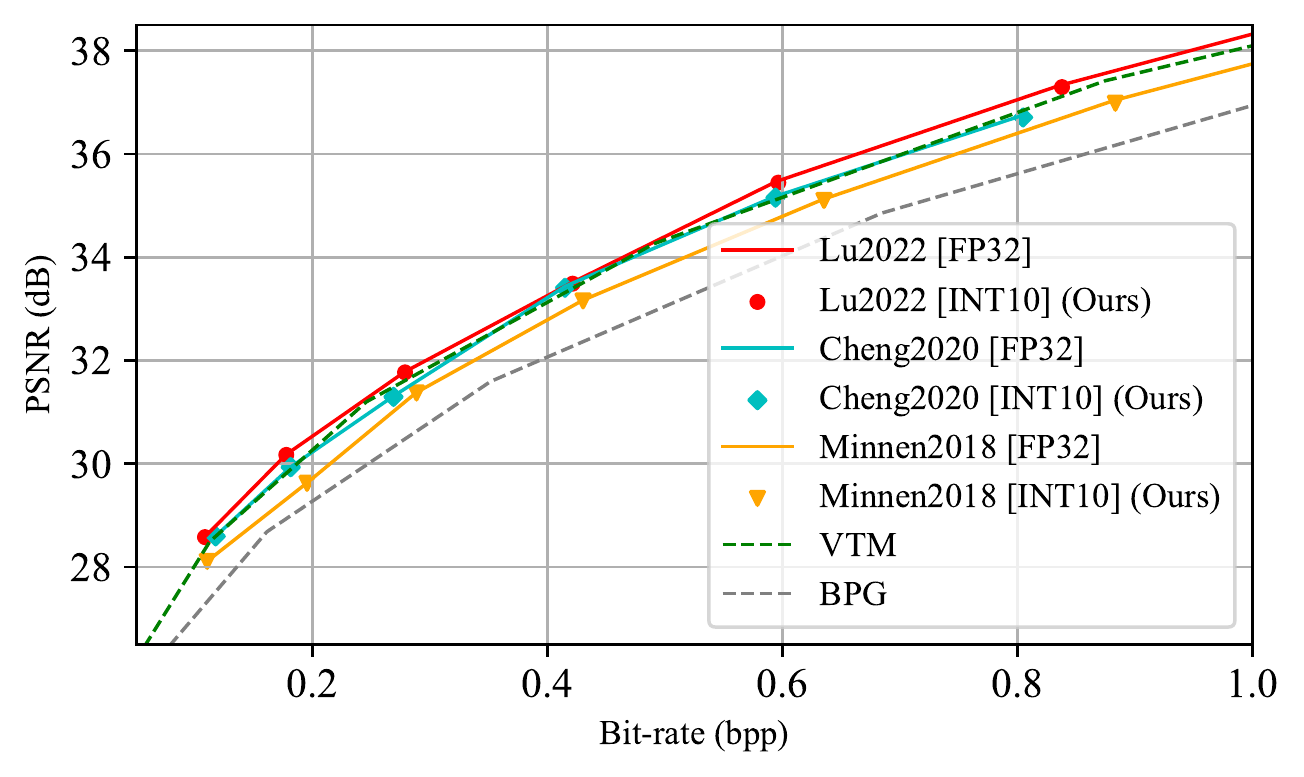} 
    }

    \subfigure[Tecnick dataset]{
        \includegraphics[width=0.42\textwidth]{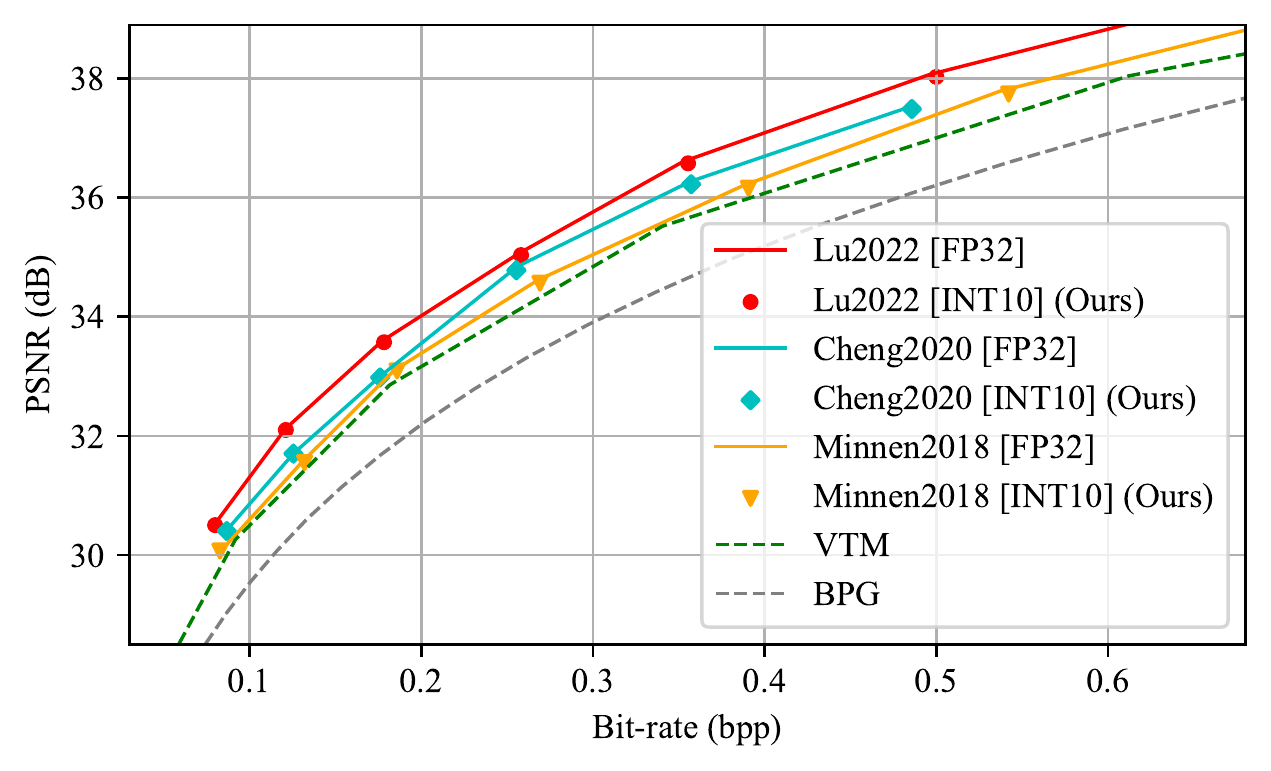} 
    }
    \caption{R-D performance of 10-bit quantization. Methods marked with [FP32] represent their original 32-bit floating-point models, while methods highlighted with [INT10] are quantized models using 10-bit fixed-point precision for processing.}
    \label{fig_appendix:RD curves}
\end{figure}

Fig. \ref{fig_appendix:RD curves} plots R-D curves for various floating-point LIC models (e.g., FP32) and their 10-bit quantized counterparts (e.g., INT10), and  Table \ref{tab_appendix:bd-rate} further reports the average R-D loss of INT10 models to their original FP32 ones using BD-Rate loss. Less than $0.5\%$ BD-Rate increase on Kodak and only about $1\%$ on Tecnick are observed, which are significantly less than the loss reported in \cite{hong2020efficient}, and also less than the loss introduced using INT8 models (see Table~\ref{tab:BD-rates}), suggesting that 10-bit quantization can mostly retain the performance of original FP32 models.

\begin{figure}[t]
\centering
\includegraphics[width=0.85\columnwidth]{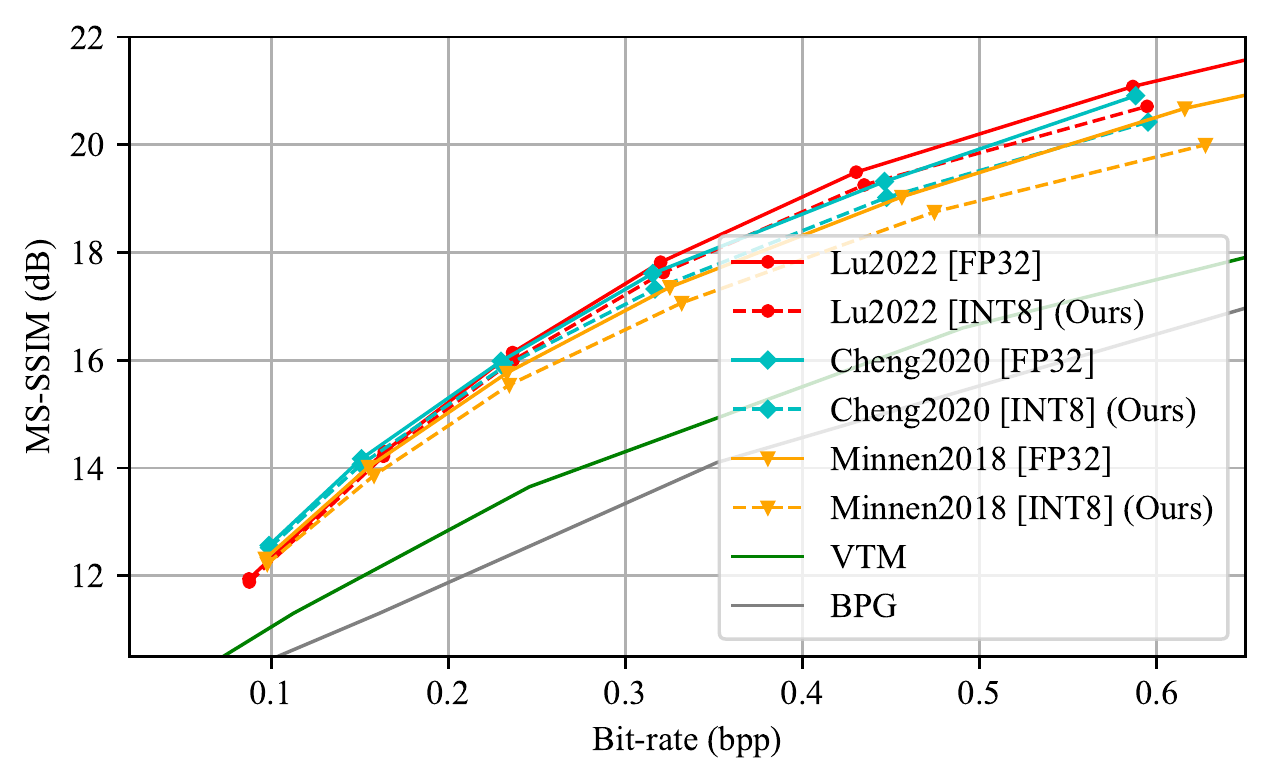}
\caption{R-D performance of MS-SSIM loss trained models on Kodak.}
\label{fig:ms-ssim}
\end{figure}

{\bf Evaluation on MS-SSIM Loss Trained Models.}
The aforementioned experiments are mainly carried out based on MSE loss trained models. We further evaluate MS-SSIM loss trained models in three testing datasets (e.g., Kodak, Tecnick, and CLIC) with three models (e.g., Lu2022 \cite{TinyLIC}, Cheng2020 \cite{cheng2020learned}, Minnen2018 \cite{minnenbt18}). For each model, six bitrates are experimented with by setting $\lambda$s  from $\{2.40,\ 4.58,\ 8.73,\ 16.64,\ 31.73,\ 60.50\}$. Two PTQ methods, FQ-ViT \cite{lin2022fqvit} and RAQ \cite{hong2020efficient}, are used for comparison.
The results are shown in Fig.~\ref{fig:ms-ssim} and Table \ref{tab:ms-ssim}. As seen, our method is also effective for MS-SSIM loss trained models.

\begin{table}[htbp]
  \centering
  \caption{BD-rate loss over various floating-point models under 8-bit quantization.}
    \begin{tabular}{ccccc}
    \toprule
    \multirow{2}[2]{*}{Model} & \multirow{2}[2]{*}{Method} & \multicolumn{3}{c}{Datasets} \\
          &       & Kodak & Tecnick & CLIC \\
    \midrule
    \multirow{3}[2]{*}{Lu2022 \cite{TinyLIC}} & FQ-ViT \cite{lin2022fqvit} & 5.97\% & 10.83\% & 9.71\% \\
          & RAQ \cite{hong2020efficient}   & 15.47\% & 23.75\% & 23.12\% \\
          & Ours  & \pmb{3.01\%} & \pmb{6.18\%} & \pmb{5.60\%} \\
    \midrule
    \multirow{2}[2]{*}{Cheng2020 \cite{cheng2020learned}} & RAQ \cite{hong2020efficient}   & 24.13\% & 29.68\% & 29.93\% \\
          & Ours  & \pmb{4.65\%} & \pmb{5.22\%} & \pmb{5.03\%} \\
    \midrule
    \multirow{2}[2]{*}{Minnen2018 \cite{minnenbt18}} & RAQ \cite{hong2020efficient}   & 31.36\% & 35.52\% & 34.97\% \\
          & Ours  & \pmb{6.15\%} & \pmb{8.83\%} & \pmb{9.16\%} \\
    \bottomrule
    \end{tabular}%
  \label{tab:ms-ssim}%
\end{table}%

\subsection{Ablations Studies}

Unless otherwise stated, all experiments are performed on the sixth bitrate ($\lambda = 0.0483$ for MSE and $\lambda = 60.50$ for MS-SSIM) of Lu2022. Other settings share similar behaviors. The default testing dataset in ablation studies is Kodak.

{\bf The initialization of quantization parameters.} We study the effect of different initialization methods on quantization parameters (e.g., scaling factors). Two main initialization methods \cite{nagel2021white}, e.g., Min-Max and grid searching, are examined. The line marked with \emph{w/o opt} means that we directly use initial quantization parameters without further RDO optimization, as in Algorithm \ref{algorithm}, while lines with \emph{w/ opt} suggest that quantization parameters are optimized with the proposed approach.

The Min-Max method determines the quantization parameters based on the maximum and minimum values of the underlying parameters. For example, the scaling factor $s$ is initiated by $s = (x_{max} - x_{min})/(2^b - 1)$, where $b$ is a predefined quantization bit-width. In grid searching, MSE loss is used as a search objective, where quantization parameter search iterates ten steps  in a given range. The results are shown in Table \ref{tab:initialization}. Both methods show close performance, but grid searching is more time-consuming. Although more search steps may lead to better performance for grid searching, the time complexity is even longer and unbearable. Thus, this work uses the Min-Max method for quantization parameters initialization.

\begin{table}[htbp]
  \centering
  \caption{The effect of Initialization methods on Kodak.}
    \begin{tabular}{ccccc}
    \toprule
          &       &       & Min-Max & grid searching \\
    \midrule
    \multicolumn{1}{c}{\multirow{4}[4]{*}{MSE}} & \multirow{2}[2]{*}{w/o opt} & PSNR  & 37.01 & 37.04 \\
          &       & Bpp   & 0.8890 & 0.8884 \\
\cmidrule{2-5}          & \multirow{2}[2]{*}{w/ opt} & PSNR  & 37.02  & 37.00 \\
          &       & Bpp   & 0.8540  & 0.8577 \\
    \midrule
    \multicolumn{1}{c}{\multirow{4}[4]{*}{MS-SSIM}} & \multirow{2}[2]{*}{w/o opt} & MS-SSIM & 20.77  & 20.80 \\
          &       & Bpp   & 0.5987  & 0.5976 \\
\cmidrule{2-5}          & \multirow{2}[2]{*}{w/ opt} & MS-SSIM & 20.74  & 20.69 \\
          &       & Bpp   & 0.5916  & 0.5914 \\
    \midrule
    \multicolumn{3}{c}{Average Time} & 14s    & 160s \\
    \bottomrule
    \end{tabular}%
  \label{tab:initialization}%
\end{table}%


{\bf Quantization Granularity.} In Fig. \ref{eq:channel_wise_weight}, we visualized the weight and activation distribution of two channels for a given layer, showing that the parameter's intensity levels are distributed differently from one channel to another. Here, we study the effect of two quantization granularities, e.g., solely layer-wise (LW) quantization and the proposed channel-wise (CW) quantization, on the performance of the final quantized model. We keep the same channel-wise activation quantization for simplicity in studying the weight quantization. As shown in Table \ref{tab:granularity}, the proposed CW method not only reduces the Bpp but also improves the PSNR, showing the R-D (rate-distortion) performance closer to the floating-point (FP32) model. The study of activation quantization granularity exhibits similar results. Therefore, we adopt channel-wise quantization granularity as our default setting in this paper.

\begin{table}[htbp]
  \centering
  \caption{Comparison of layer-wise (LW) and channel-wise (CW) weight quantization.}
    \begin{tabular}{cccccc}
    \toprule
          &       & \multicolumn{2}{c}{MSE models} & \multicolumn{2}{c}{MS-SSIM models} \\
          &       & PSNR  & Bpp   & MS-SSIM & Bpp \\
    \midrule
          & FP32  & 37.33 & 0.8354 & 21.08 & 0.5866 \\
    \midrule
     & LW    & 36.24 & 0.8658 & 20.48 & 0.5966 \\
          & \pmb{CW}    & \pmb{37.02} & \pmb{0.8540} & \pmb{20.74} & \pmb{0.591} \\
    \bottomrule
    \end{tabular}%
  \label{tab:granularity}%
\end{table}%

\begin{figure}[htbp]
    \centering
    \includegraphics[width=0.8\columnwidth]{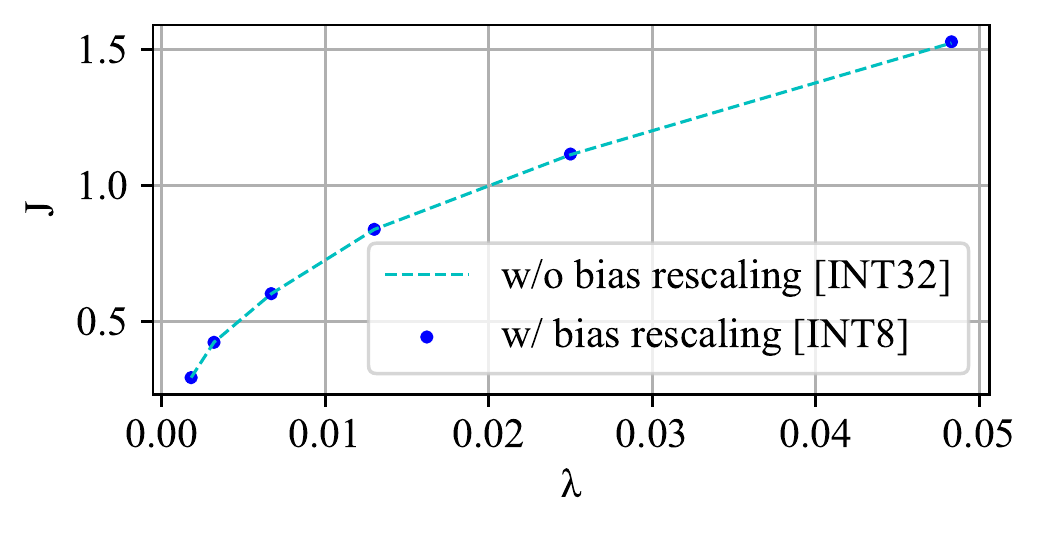}
    \caption{Effect of bias rescaling. We choose 32-bit bias with Min-Max quantization as "Baseline".}
    \label{fig:bias}
\end{figure}

{\bf Bias Rescaling.}
Taking INT32 bias without rescaling as the reference, we test the effect of bias rescaling on Lu2022. The common Min-Max quantization is adopted without quantizing activation, which does not require the calibration set and does not introduce quantization error of activation, which generally simplifies the problem for discussion. As shown in Fig. \ref{fig:bias}, bias rescaling for fully 8-bit processing has no negative impact on performance. 

\begin{figure}[htbp]
    \centering
    \includegraphics[width=0.75\columnwidth]{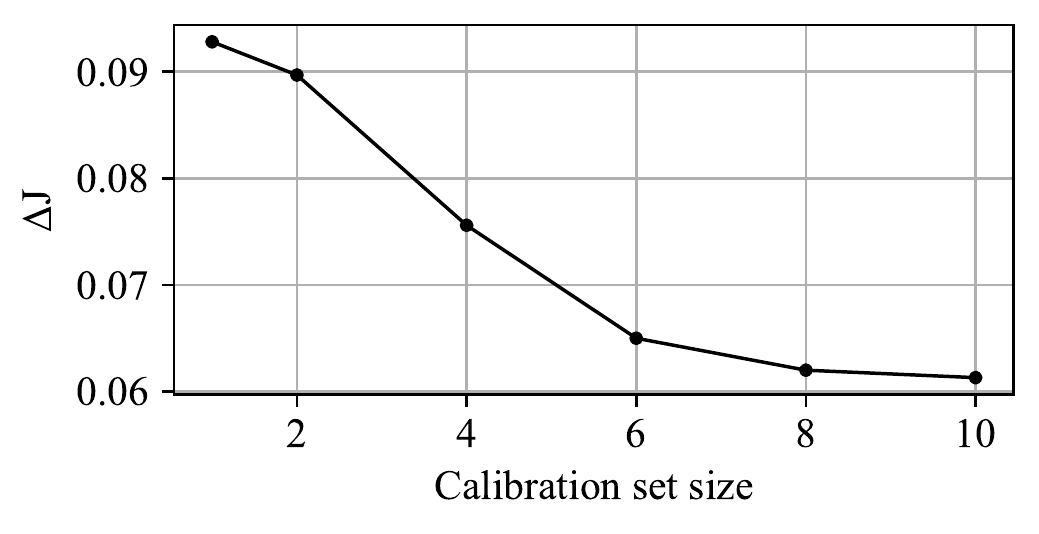}
    \caption{The variance of RD-loss as calibration set size increases.}
    \label{set_size}
\end{figure}

{\bf The Size of Calibration Set.}
We randomly select images from ImageNet \cite{russakovsky2015imagenet} to formulate the calibration set for quantization parameters determination in our RDO-PTQ. As shown in Fig. \ref{set_size}, when the size of the calibration set increases to 10, there is basically no loss to the R-D performance compared with the floating point model. Therefore, we adopt 10 images for calibration in this work.

\begin{figure}[t]
    \centering
    \subfigure[MSE]{
        \includegraphics[width=0.46\columnwidth]{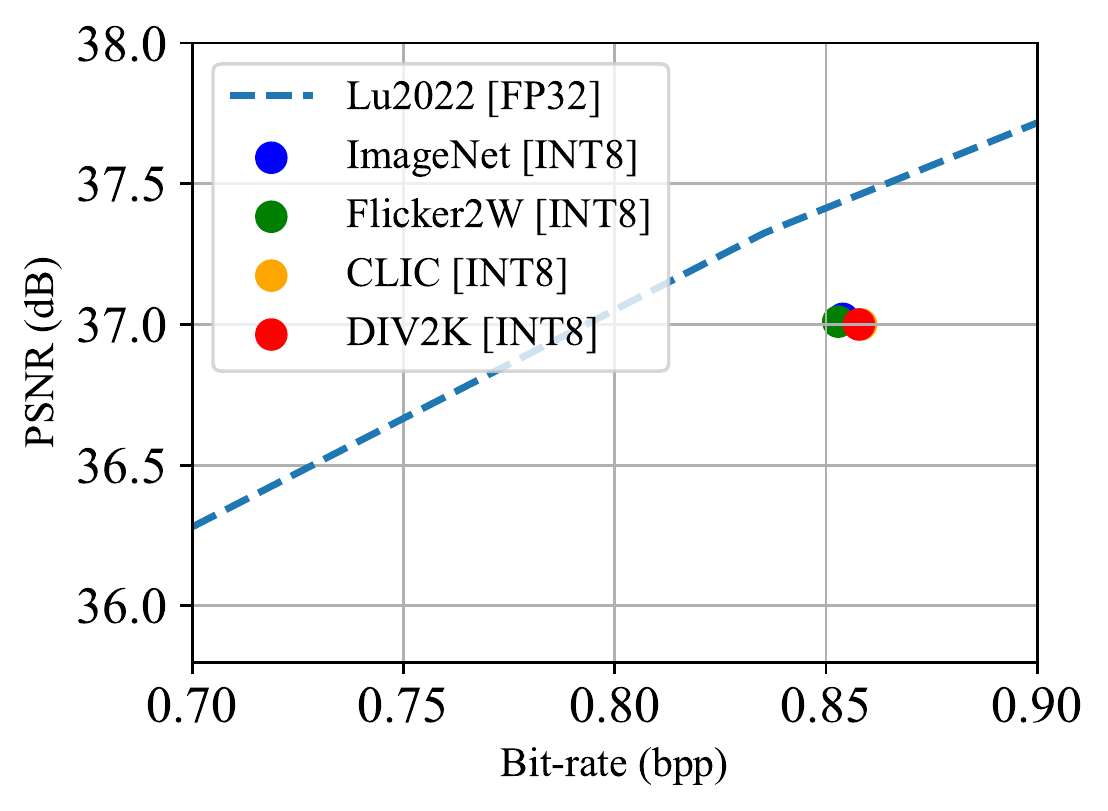} 
        \label{fig:mse_c}
    }
    \subfigure[MS-SSIM]{
        \includegraphics[width=0.46\columnwidth]{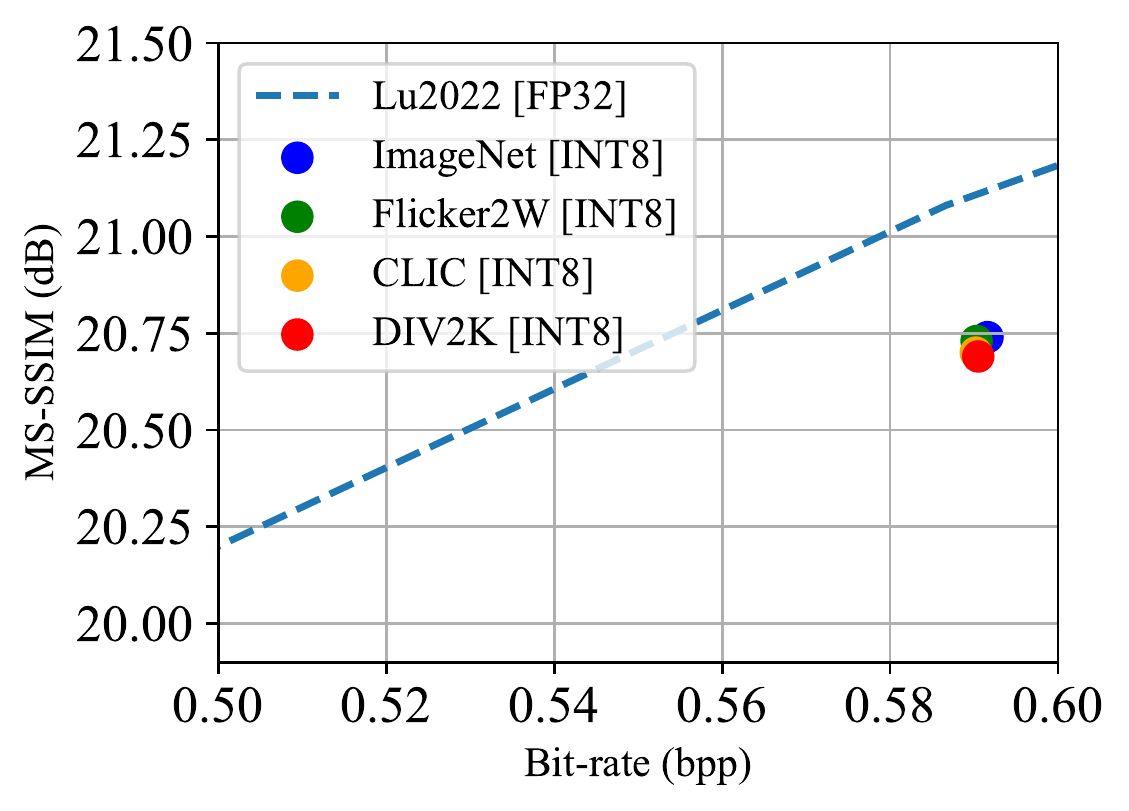}
        \label{fig:ms-ssim_c}
    }
    \caption{The effect of using different calibration set to optimize model quantization parameters. Kodak images are tested.}
    \label{fig:calidata}
\end{figure}

{\bf The Selection of Calibration Set.}
The proposed calibration dataset is a tiny-scale dataset with very few images for optimizing model quantization parameters. Any popular image-based dataset, including the training set, can form our calibration dataset by randomly selecting image samples, as long as it does not overlap with the testing dataset. To evaluate the impact of different calibration datasets on the performance of the quantized model, we examine four different calibration datasets formulated using ImageNet \cite{russakovsky2015imagenet}, Flicker2W \cite{flicker2w}, CLIC, and DIV2K \cite{DIV2K}, respectively.
Note that Flicker2W \cite{flicker2w} is the training dataset used in Lu2022. As shown in Fig. \ref{fig:calidata}, the impact of the calibration set on the results is negligible, with their performance measures noticeably overlapped.

\section{Generalization Ability}
RDO-PTQ is a task-loss-optimized  PTQ scheme.  We subsequently show that such a task-loss-optimized PTQ not only works for the learned image compression but also point cloud compression (see details in supplementary material), super-resolution (SR), and image classification tasks. SR's task loss is PSNR  between the scaled image and its corresponding label, while the classification uses Cross-Entropy as task loss. We respectively adjust the corresponding loss function to perform task-loss-optimized PTQ following the same procedures in Algorithm~\ref{algorithm}.

\subsection{Super-Resolution (SR)}
{\bf Experimental Settings.}
We choose two widely-used SR models, namely EDSR \cite{lim2017enhanced} and SwinIR \cite{liang2021swinir}, for quantization. For instance, EDSR relies on stacked convolutions while SwinIR adopts self-attention processing, both of which demonstrate outstanding performance. We use their publicly available pre-trained models and apply the proposed quantization method without modifying the SR network. 

For testing, we use four standard benchmark datasets: Set5 \cite{set5}, Set14 \cite{set14}, BSD100 \cite{B100}, and Urban100 \cite{urban100}. For the evaluation metrics, we use the PSNR between the Y channel of the output image and its original high-resolution label. 

{\bf Evaluation.}
For EDSR \cite{lim2017enhanced}, we quantize the weights, bias, and activation of all layers into the 8-bit. The PAMS \cite{li2020pams} is compared where its data is directly obtained from the publicly available source. As shown in Table \ref{tab:sr}, the proposed method with 8-bit quantization achieves comparable performance to the original floating-point model across all different datasets and scaling ratios, e.g., reporting 0.01-0.02dB PSNR loss and demonstrates better efficiency than state-of-the-art PAMS noticeably. 

For SwinIR \cite{liang2021swinir}, we adopt its lightweight models as the backbone and use 8-bit quantization. In experiments, we observe that the final output of the Residual Swin Transformer Block (RSTB) exhibits a long-tailed distribution, which mainly contributes to the performance degradation of the quantization. Thus, the final layer of RSTB is quantized to 10-bit instead while keeping other layers to 8-bit precision for a justified trade-off between performance and bit-width.
To date, we are unaware of published works regarding the SwinIR quantization; therefore, we do not compare our method with other approaches. As shown in Table \ref{tab:sr}, the proposed method can quantize models with negligible performance loss.

\begin{table}[htbp]

    \setlength\tabcolsep{5pt} 
  \centering
  \caption{Comparison of existing methods on EDSR \cite{lim2017enhanced} and SwinIR \cite{liang2021swinir} with different scales. }
    \begin{tabular}{ccccccc}
    \toprule
    \multirow{2}[2]{*}{Method} & \multirow{2}[2]{*}{Scale} & Bit-width & \multicolumn{4}{c}{PSNR (dB)} \\
          &       & w/a   & Set5  & Set14 & B100  & Urban100 \\
    \midrule
    EDSR  & \multirow{3}[2]{*}{x2} & 32/32 & 38.085 & 33.678 & 32.206 & 32.156 \\
    PAMS  &       & 8/8 & 37.946 & 33.564 & 32.157 & 32.003 \\
    Ours  &       & 8/8 & \pmb{38.065} & \pmb{33.669} & \pmb{32.199} & \pmb{32.141} \\
    \midrule
    EDSR  & \multirow{3}[2]{*}{x3} & 32/32 & 34.523 & 30.402 & 29.157 & 28.323 \\
    PAMS  &       & 8/8 & 34.361 & 30.280 & 29.079 & 28.142 \\
    Ours  &       & 8/8 & \pmb{34.517} & \pmb{30.397} & \pmb{29.155} & \pmb{28.314} \\
    \midrule
    EDSR  & \multirow{3}[2]{*}{x4} & 32/32 & 32.288 & 28.670 & 27.629 & 26.188 \\
    PAMS  &       & 8/8 & 32.124 & 28.585 & 27.565 & 26.016 \\
    Ours  &       & 8/8 & \pmb{32.281} & \pmb{28.666} & \pmb{27.628} & \pmb{26.188} \\
    \midrule
    SwinIR & \multirow{2}[2]{*}{x2} & 32/32 & 38.139 & 33.857 & 32.306 & 32.759 \\
    Ours  &       & 8/8 & \pmb{38.063} & \pmb{33.821} & \pmb{32.279} & \pmb{32.710} \\
    \midrule
    SwinIR & \multirow{2}[2]{*}{x3} & 32/32 & 34.621 & 30.538 & 29.203 & 28.656 \\
    Ours  &       & 8/8 & \pmb{34.591} & \pmb{30.518} & \pmb{29.192} & \pmb{28.638} \\
    \midrule
    SwinIR & \multirow{2}[2]{*}{x4} & 32/32 & 32.442 & 28.768 & 27.689 & 26.474 \\
    Ours  &       & 8/8 & \pmb{32.419} & \pmb{28.754} & \pmb{27.673} & \pmb{26.462} \\
    \bottomrule
    \end{tabular}%
  \label{tab:sr}%
\end{table}%

\subsection{Image Classification}
{\bf Experimental Settings.} We conduct experiments on two popular architectures, including ResNet-18 \cite{resnet} with normal stacked convolution and RegNet-600MF \cite{regnet} with group convolution. For testing, the ILSVRC-2012ImageNet \cite{russakovsky2015imagenet} (we refer to it as ImageNet) dataset is utilized to evaluate the quantization performance. Such an ImageNet dataset contains 1.2 million training images and 50K validation images labeled for 1,000 categories.

We compare our method with multiple strong baselines, including RAQ \cite{hong2020efficient}, AdaRound \cite{nagel2020up}, and BRECQ \cite{li2020brecq}. w and a indicate the bit-width of the weight and activation, respectively.
Following their algorithm description, we faithfully reproduce AdaRound and BRECQ, where our implementation reveals the similar performance reported in their original papers.




\begin{table}[htbp]
  \centering
  \caption{Classification comparison of existing methods on ResNet \cite{resnet} and RegNet \cite{regnet} with ImageNet.}
    \begin{tabular}{ccccc}
    \toprule
    \multirow{2}[2]{*}{Method} & \multicolumn{2}{c}{Bit-width} & \multirow{2}[2]{*}{ResNet-18} & \multirow{2}[2]{*}{RegNet-600MF} \\
          & w     & a     &       &  \\
    \midrule
    FP32  & 32    & 32    & 71.01\% & 73.54\% \\
    RAQ \cite{hong2020efficient}*   & 8     & 8     & 70.78\% & 73.41\% \\
    AdaRound \cite{nagel2020up}* & 8     & 8     & 70.93\% & 73.48\% \\
    BRECQ \cite{li2020brecq}* & 8     & 8     & 70.94\% & 73.46\% \\
    Ours  & 8     & 8     & \pmb{70.96\%} & \pmb{73.50\%} \\
    \midrule
    AdaRound \cite{nagel2020up}* & 4     & 4     & 68.06\% & 68.50\% \\
    BRECQ \cite{li2020brecq}* & 4     & 4     & 68.14\% & 68.03\% \\
    Ours  & 4     & 4     & \pmb{68.83\%} & \pmb{70.46\%} \\
    \midrule
    AdaRound \cite{nagel2020up}* & 2     & 2     & 15.40\% & 1.40\% \\
    BRECQ \cite{li2020brecq}* & 2     & 2     & 31.34\% & 1.17\% \\
    Ours  & 2     & 2     & \pmb{42.65\%} & \pmb{31.52\%} \\
    \bottomrule
    \end{tabular}%
  \label{tab:ic}%
\end{table}%

{\bf Evaluation.} The results are summarized in Table \ref{tab:ic}. As seen, our method outperforms other PTQ baselines. First, in the W8A8 quantization setting, our method only suffers $0.05\%$ and $0.04\%$ accuracy loss, respectively, on ResNet-18 and RegNet-600MF. At the same time, the storage overhead of the model is reduced to 1/4 of the original floating-point model. Here, the gap between our method and AdaRound/BRECQ is negligible. However, the gap becomes more apparent as the bit-width decreases. When applying the w4a4 quantization setting, our method can achieve $0.5\% - 2.5\%$ accuracy improvement than AdaRound and BRECQ. Such an accuracy gap becomes significant when using w2a2, offering over $10\%$ and $30\%$ accuracy improvements for ResNet-18 and RegNet-600MF, respectively. 


\section{Conclusion} \label{sec:conclusion}
This paper studied the PTQ to directly quantize off-the-shelf, pretrained floating-point image compression models for their fixed-point counterparts. To retain the compression efficiency as the native floating-point LICs, we suggested the R-D optimized PTQ, which was first justified theoretically and then proved experimentally. In RDO-PTQ, we determined channel-wise weight and activation ranges from one layer to another, and layer-wise bias ranges for subsequent rescaling, to which a tiny set of images (e.g., about ten samples in total) were sufficient to optimize quantization parameters. Results revealed the superior efficiency of the proposed method, presenting large performance improvements over existing approaches. More importantly, {our method did not require retraining model parameters but only adjusting quantization parameters and offered a push-button solution.} An interesting topic for further study is reducing the quantization loss at high bitrates. Besides, we extended our method to the super-resolution and image classification tasks. All of them have revealed our method is reliable and robust to provide comparable performance to the original floating-point models.

\bibliography{arxiv_v3}

\begin{thebibliography}{10}
\providecommand{\url}[1]{#1}
\csname url@samestyle\endcsname
\providecommand{\newblock}{\relax}
\providecommand{\bibinfo}[2]{#2}
\providecommand{\BIBentrySTDinterwordspacing}{\spaceskip=0pt\relax}
\providecommand{\BIBentryALTinterwordstretchfactor}{4}
\providecommand{\BIBentryALTinterwordspacing}{\spaceskip=\fontdimen2\font plus
\BIBentryALTinterwordstretchfactor\fontdimen3\font minus
  \fontdimen4\font\relax}
\providecommand{\BIBforeignlanguage}[2]{{%
\expandafter\ifx\csname l@#1\endcsname\relax
\typeout{** WARNING: IEEEtran.bst: No hyphenation pattern has been}%
\typeout{** loaded for the language `#1'. Using the pattern for}%
\typeout{** the default language instead.}%
\else
\language=\csname l@#1\endcsname
\fi
#2}}
\providecommand{\BIBdecl}{\relax}
\BIBdecl

\bibitem{JPEG}
G.~Wallace, ``The jpeg still picture compression standard,'' \emph{IEEE
  Transactions on Consumer Electronics}, vol.~38, no.~1, pp. xviii--xxxiv,
  1992.

\bibitem{BPG}
G.~J. Sullivan, J.-R. Ohm, W.-J. Han, and T.~Wiegand, ``Overview of the high
  efficiency video coding (hevc) standard,'' \emph{IEEE Transactions on
  Circuits and Systems for Video Technology}, vol.~22, no.~12, pp. 1649--1668,
  2012.

\bibitem{balle2016end}
J.~Ball{\'e}, V.~Laparra, and E.~P. Simoncelli, ``End-to-end optimized image
  compression,'' in \emph{5th International Conference on Learning
  Representations, ICLR 2017}, 2017.

\bibitem{chen2017deepcoder}
T.~Chen, H.~Liu, Q.~Shen, T.~Yue, X.~Cao, and Z.~Ma, ``Deepcoder: A deep neural
  network based video compression,'' in \emph{2017 IEEE Visual Communications
  and Image Processing (VCIP)}.\hskip 1em plus 0.5em minus 0.4em\relax IEEE,
  2017, pp. 1--4.

\bibitem{jacob2018quantization}
B.~Jacob, S.~Kligys, B.~Chen, M.~Zhu, M.~Tang, A.~Howard, H.~Adam, and
  D.~Kalenichenko, ``Quantization and training of neural networks for efficient
  integer-arithmetic-only inference,'' in \emph{Proceedings of the IEEE
  conference on computer vision and pattern recognition}, 2018, pp. 2704--2713.

\bibitem{VVC}
B.~Bross, Y.-K. Wang, Y.~Ye, S.~Liu, J.~Chen, G.~J. Sullivan, and J.-R. Ohm,
  ``Overview of the versatile video coding (vvc) standard and its
  applications,'' \emph{IEEE Transactions on Circuits and Systems for Video
  Technology}, vol.~31, no.~10, pp. 3736--3764, 2021.

\bibitem{he2022PTQ}
D.~He, Z.~Yang, Y.~Chen, Q.~Zhang, H.~Qin, and Y.~Wang, ``Post-training
  quantization for cross-platform learned image compression,'' \emph{arXiv
  preprint arXiv:2202.07513}, 2022.

\bibitem{dai2021vs}
S.~Dai, R.~Venkatesan, M.~Ren, B.~Zimmer, W.~Dally, and B.~Khailany,
  ``Vs-quant: Per-vector scaled quantization for accurate low-precision neural
  network inference,'' \emph{Proceedings of Machine Learning and Systems},
  vol.~3, pp. 873--884, 2021.

\bibitem{nagel2020up}
M.~Nagel, R.~A. Amjad, M.~Van~Baalen, C.~Louizos, and T.~Blankevoort, ``Up or
  down? adaptive rounding for post-training quantization,'' in
  \emph{International Conference on Machine Learning}.\hskip 1em plus 0.5em
  minus 0.4em\relax PMLR, 2020, pp. 7197--7206.

\bibitem{sun2021learned}
H.~Sun, L.~Yu, and J.~Katto, ``Learned image compression with fixed-point
  arithmetic,'' in \emph{2021 Picture Coding Symposium (PCS)}.\hskip 1em plus
  0.5em minus 0.4em\relax IEEE, 2021, pp. 1--5.

\bibitem{bhalgat2020lsq+}
Y.~Bhalgat, J.~Lee, M.~Nagel, T.~Blankevoort, and N.~Kwak, ``Lsq+: Improving
  low-bit quantization through learnable offsets and better initialization,''
  in \emph{Proceedings of the IEEE/CVF Conference on Computer Vision and
  Pattern Recognition Workshops}, 2020, pp. 696--697.

\bibitem{carion2020end}
N.~Carion, F.~Massa, G.~Synnaeve, N.~Usunier, A.~Kirillov, and S.~Zagoruyko,
  ``End-to-end object detection with transformers,'' in \emph{European
  conference on computer vision}.\hskip 1em plus 0.5em minus 0.4em\relax
  Springer, 2020, pp. 213--229.

\bibitem{liu2021swin}
Z.~Liu, Y.~Lin, Y.~Cao, H.~Hu, Y.~Wei, Z.~Zhang, S.~Lin, and B.~Guo, ``Swin
  transformer: Hierarchical vision transformer using shifted windows,'' in
  \emph{Proceedings of the IEEE/CVF International Conference on Computer
  Vision}, 2021, pp. 10\,012--10\,022.

\bibitem{TIC}
M.~Lu, P.~Guo, H.~Shi, C.~Cao, and Z.~Ma, ``Transformer-based image
  compression,'' in \emph{2022 Data Compression Conference (DCC)}, 2022, pp.
  469--469.

\bibitem{TinyLIC}
M.~Lu and Z.~Ma, ``High-efficiency lossy image coding through adaptive
  neighborhood information aggregation,'' \emph{arXiv preprint
  arXiv:2204.11448}, 2022.

\bibitem{hubara2020improving}
I.~Hubara, Y.~Nahshan, Y.~Hanani, R.~Banner, and D.~Soudry, ``Improving post
  training neural quantization: Layer-wise calibration and integer
  programming,'' \emph{arXiv preprint arXiv:2006.10518}, 2020.

\bibitem{ballemshj18}
J.~Ball{\'{e}}, D.~Minnen, S.~Singh, S.~J. Hwang, and N.~Johnston,
  ``Variational image compression with a scale hyperprior,'' in \emph{6th
  International Conference on Learning Representations, {ICLR} 2018, Vancouver,
  BC, Canada, April 30 - May 3, 2018, Conference Track Proceedings}.\hskip 1em
  plus 0.5em minus 0.4em\relax OpenReview.net, 2018.

\bibitem{xie2021enhanced}
Y.~Xie, K.~L. Cheng, and Q.~Chen, ``Enhanced invertible encoding for learned
  image compression,'' in \emph{Proceedings of the 29th ACM International
  Conference on Multimedia}, 2021, pp. 162--170.

\bibitem{qian2020learning}
Y.~Qian, Z.~Tan, X.~Sun, M.~Lin, D.~Li, Z.~Sun, L.~Hao, and R.~Jin, ``Learning
  accurate entropy model with global reference for image compression,'' in
  \emph{International Conference on Learning Representations}, 2020.

\bibitem{kim2022joint}
J.-H. Kim, B.~Heo, and J.-S. Lee, ``Joint global and local hierarchical priors
  for learned image compression,'' in \emph{Proceedings of the IEEE/CVF
  Conference on Computer Vision and Pattern Recognition}, 2022, pp. 5992--6001.

\bibitem{cheng2020learned}
Z.~Cheng, H.~Sun, M.~Takeuchi, and J.~Katto, ``Learned image compression with
  discretized gaussian mixture likelihoods and attention modules,'' in
  \emph{Proceedings of the IEEE/CVF Conference on Computer Vision and Pattern
  Recognition}, 2020, pp. 7939--7948.

\bibitem{minnenbt18}
D.~Minnen, J.~Ball{\'{e}}, and G.~Toderici, ``Joint autoregressive and
  hierarchical priors for learned image compression,'' in \emph{Advances in
  Neural Information Processing Systems 31: Annual Conference on Neural
  Information Processing Systems 2018, NeurIPS 2018, 3-8 December 2018,
  Montr{\'{e}}al, Canada}, S.~Bengio, H.~M. Wallach, H.~Larochelle, K.~Grauman,
  N.~Cesa{-}Bianchi, and R.~Garnett, Eds., 2018, pp. 10\,794--10\,803.

\bibitem{zhu2021transformer}
Y.~Zhu, Y.~Yang, and T.~Cohen, ``Transformer-based transform coding,'' in
  \emph{International Conference on Learning Representations}, 2021.

\bibitem{minnen2020channel}
D.~Minnen and S.~Singh, ``Channel-wise autoregressive entropy models for
  learned image compression,'' in \emph{2020 IEEE International Conference on
  Image Processing (ICIP)}.\hskip 1em plus 0.5em minus 0.4em\relax IEEE, 2020,
  pp. 3339--3343.

\bibitem{balle2018integer}
J.~Ball{\'e}, N.~Johnston, and D.~Minnen, ``Integer networks for data
  compression with latent-variable models,'' in \emph{International Conference
  on Learning Representations}, 2018.

\bibitem{sun2020end}
H.~Sun, Z.~Cheng, M.~Takeuchi, and J.~Katto, ``End-to-end learned image
  compression with fixed point weight quantization,'' in \emph{2020 IEEE
  International Conference on Image Processing (ICIP)}.\hskip 1em plus 0.5em
  minus 0.4em\relax IEEE, 2020, pp. 3359--3363.

\bibitem{hong2020efficient}
W.~Hong, T.~Chen, M.~Lu, S.~Pu, and Z.~Ma, ``Efficient neural image decoding
  via fixed-point inference,'' \emph{IEEE Transactions on Circuits and Systems
  for Video Technology}, vol.~31, no.~9, pp. 3618--3630, 2020.

\bibitem{Rate-distortion-Theory}
L.~Davisson, ``Rate distortion theory: A mathematical basis for data
  compression,'' \emph{IEEE Transactions on Communications}, vol.~20, no.~6,
  pp. 1202--1202, 1972.

\bibitem{Low-Bit-PTQ}
Y.~Choukroun, E.~Kravchik, F.~Yang, and P.~Kisilev, ``Low-bit quantization of
  neural networks for efficient inference,'' in \emph{2019 IEEE/CVF
  International Conference on Computer Vision Workshop (ICCVW)}, 2019, pp.
  3009--3018.

\bibitem{hu2021learning}
Y.~Hu, W.~Yang, Z.~Ma, and J.~Liu, ``Learning end-to-end lossy image
  compression: A benchmark,'' \emph{IEEE Transactions on Pattern Analysis and
  Machine Intelligence}, 2021.

\bibitem{chen2021end}
T.~Chen, H.~Liu, Z.~Ma, Q.~Shen, X.~Cao, and Y.~Wang, ``End-to-end learnt image
  compression via non-local attention optimization and improved context
  modeling,'' \emph{IEEE Transactions on Image Processing}, vol.~30, pp.
  3179--3191, 2021.

\bibitem{sun2022q}
H.~Sun, L.~Yu, and J.~Katto, ``Q-lic: Quantizing learned image compression with
  channel splitting,'' \emph{IEEE Transactions on Circuits and Systems for
  Video Technology}, 2022.

\bibitem{nagel2021white}
M.~Nagel, M.~Fournarakis, R.~A. Amjad, Y.~Bondarenko, M.~van Baalen, and
  T.~Blankevoort, ``A white paper on neural network quantization,'' \emph{arXiv
  preprint arXiv:2106.08295}, 2021.

\bibitem{zhao2021efficient}
H.~Zhao, D.~Liu, and H.~Li, ``Efficient integer-arithmetic-only convolutional
  networks with bounded relu,'' in \emph{2021 IEEE International Symposium on
  Circuits and Systems (ISCAS)}.\hskip 1em plus 0.5em minus 0.4em\relax IEEE,
  2021, pp. 1--5.

\bibitem{li2020pams}
H.~Li, C.~Yan, S.~Lin, X.~Zheng, B.~Zhang, F.~Yang, and R.~Ji, ``Pams:
  Quantized super-resolution via parameterized max scale,'' in \emph{European
  Conference on Computer Vision}.\hskip 1em plus 0.5em minus 0.4em\relax
  Springer, 2020, pp. 564--580.

\bibitem{botev2017practical}
A.~Botev, H.~Ritter, and D.~Barber, ``Practical gauss-newton optimisation for
  deep learning,'' in \emph{International Conference on Machine
  Learning}.\hskip 1em plus 0.5em minus 0.4em\relax PMLR, 2017, pp. 557--565.

\bibitem{gibson2017rate}
J.~Gibson, ``Rate distortion functions and rate distortion function lower
  bounds for real-world sources,'' \emph{Entropy}, vol.~19, no.~11, p. 604,
  2017.

\bibitem{kochenberger2014unconstrained}
G.~Kochenberger, J.-K. Hao, F.~Glover, M.~Lewis, Z.~L{\"u}, H.~Wang, and
  Y.~Wang, ``The unconstrained binary quadratic programming problem: a
  survey,'' \emph{Journal of combinatorial optimization}, vol.~28, no.~1, pp.
  58--81, 2014.

\bibitem{begaint2020compressai}
J.~B{\'e}gaint, F.~Racap{\'e}, S.~Feltman, and A.~Pushparaja, ``Compressai: a
  pytorch library and evaluation platform for end-to-end compression
  research,'' \emph{arXiv preprint arXiv:2011.03029}, 2020.

\bibitem{lin2022fqvit}
Y.~Lin, T.~Zhang, P.~Sun, Z.~Li, and S.~Zhou, ``Fq-vit: Post-training
  quantization for fully quantized vision transformer,'' in \emph{Proceedings
  of the Thirty-First International Joint Conference on Artificial
  Intelligence, {IJCAI-22}}, 2022, pp. 1173--1179.

\bibitem{liu2021post}
Z.~Liu, Y.~Wang, K.~Han, W.~Zhang, S.~Ma, and W.~Gao, ``Post-training
  quantization for vision transformer,'' \emph{Advances in Neural Information
  Processing Systems}, vol.~34, pp. 28\,092--28\,103, 2021.

\bibitem{qian2021entroformer}
Y.~Qian, X.~Sun, M.~Lin, Z.~Tan, and R.~Jin, ``Entroformer: A transformer-based
  entropy model for learned image compression,'' in \emph{International
  Conference on Learning Representations}, 2021.

\bibitem{bjontegaard2001calculation}
G.~Bjontegaard, ``Calculation of average psnr differences between rd-curves,''
  \emph{VCEG-M33}, 2001.

\bibitem{le2022mobilecodec}
H.~Le, L.~Zhang, A.~Said, G.~Sautiere, Y.~Yang, P.~Shrestha, F.~Yin,
  R.~Pourreza, and A.~Wiggers, ``Mobilecodec: neural inter-frame video
  compression on mobile devices,'' in \emph{Proceedings of the 13th ACM
  Multimedia Systems Conference}, 2022, pp. 324--330.

\bibitem{li2020brecq}
Y.~Li, R.~Gong, X.~Tan, Y.~Yang, P.~Hu, Q.~Zhang, F.~Yu, W.~Wang, and S.~Gu,
  ``Brecq: Pushing the limit of post-training quantization by block
  reconstruction,'' in \emph{International Conference on Learning
  Representations}, 2020.

\bibitem{choi2018pact}
J.~Choi, Z.~Wang, S.~Venkataramani, P.~I.-J. Chuang, V.~Srinivasan, and
  K.~Gopalakrishnan, ``Pact: Parameterized clipping activation for quantized
  neural networks,'' \emph{arXiv preprint arXiv:1805.06085}, 2018.

\bibitem{hinton2012neural}
G.~Hinton, N.~Srivastava, and K.~Swersky, ``Neural networks for machine
  learning,'' \emph{Coursera, video lectures}, vol. 264, no.~1, pp. 2146--2153,
  2012.

\bibitem{kingma2014adam}
D.~P. Kingma and J.~Ba, ``Adam: A method for stochastic optimization,''
  \emph{arXiv preprint arXiv:1412.6980}, 2014.

\bibitem{krishnamoorthi2018quantizing}
R.~Krishnamoorthi, ``Quantizing deep convolutional networks for efficient
  inference: A whitepaper,'' \emph{arXiv preprint arXiv:1806.08342}, 2018.

\bibitem{wu2020integer}
H.~Wu, P.~Judd, X.~Zhang, M.~Isaev, and P.~Micikevicius, ``Integer quantization
  for deep learning inference: Principles and empirical evaluation,''
  \emph{arXiv preprint arXiv:2004.09602}, 2020.

\bibitem{2020xilinx}
Xilinx, ``Convolutional neural network with int4 optimization on xilinx,'' in
  \emph{WP521 (v1.0.1) June 24}, 2020.

\bibitem{DIV2K}
R.~Timofte, E.~Agustsson, L.~Van~Gool, M.-H. Yang, and L.~Zhang, ``Ntire 2017
  challenge on single image super-resolution: Methods and results,'' in
  \emph{Proceedings of the IEEE conference on computer vision and pattern
  recognition workshops}, 2017, pp. 114--125.

\bibitem{lim2017enhanced}
B.~Lim, S.~Son, H.~Kim, S.~Nah, and K.~Mu~Lee, ``Enhanced deep residual
  networks for single image super-resolution,'' in \emph{Proceedings of the
  IEEE conference on computer vision and pattern recognition workshops}, 2017,
  pp. 136--144.

\bibitem{liang2021swinir}
J.~Liang, J.~Cao, G.~Sun, K.~Zhang, L.~Van~Gool, and R.~Timofte, ``Swinir:
  Image restoration using swin transformer,'' in \emph{Proceedings of the
  IEEE/CVF international conference on computer vision}, 2021, pp. 1833--1844.

\bibitem{set5}
M.~Bevilacqua, A.~Roumy, C.~Guillemot, and M.-L.~A. Morel, ``Low-complexity
  single-image super-resolution based on nonnegative neighbor embedding,'' in
  \emph{British Machine Vision Conference (BMVC)}, 2012.

\bibitem{set14}
C.~Ledig, L.~Theis, F.~Husz{\'a}r, J.~Caballero, A.~Cunningham, A.~Acosta,
  A.~Aitken, A.~Tejani, J.~Totz, Z.~Wang \emph{et~al.}, ``Photo-realistic
  single image super-resolution using a generative adversarial network,'' in
  \emph{Proceedings of the IEEE conference on computer vision and pattern
  recognition}, 2017, pp. 4681--4690.

\bibitem{B100}
D.~Martin, C.~Fowlkes, D.~Tal, and J.~Malik, ``A database of human segmented
  natural images and its application to evaluating segmentation algorithms and
  measuring ecological statistics,'' in \emph{Proceedings Eighth IEEE
  International Conference on Computer Vision. ICCV 2001}, vol.~2.\hskip 1em
  plus 0.5em minus 0.4em\relax IEEE, 2001, pp. 416--423.

\bibitem{urban100}
J.-B. Huang, A.~Singh, and N.~Ahuja, ``Single image super-resolution from
  transformed self-exemplars,'' in \emph{Proceedings of the IEEE conference on
  computer vision and pattern recognition}, 2015, pp. 5197--5206.

\bibitem{resnet}
K.~He, X.~Zhang, S.~Ren, and J.~Sun, ``Deep residual learning for image
  recognition,'' in \emph{Proceedings of the IEEE conference on computer vision
  and pattern recognition}, 2016, pp. 770--778.

\bibitem{regnet}
I.~Radosavovic, R.~P. Kosaraju, R.~Girshick, K.~He, and P.~Doll{\'a}r,
  ``Designing network design spaces,'' in \emph{Proceedings of the IEEE/CVF
  conference on computer vision and pattern recognition}, 2020, pp.
  10\,428--10\,436.

\bibitem{hong2022daq}
C.~Hong, H.~Kim, S.~Baik, J.~Oh, and K.~M. Lee, ``Daq: Channel-wise
  distribution-aware quantization for deep image super-resolution networks,''
  in \emph{Proceedings of the IEEE/CVF Winter Conference on Applications of
  Computer Vision}, 2022, pp. 2675--2684.

\bibitem{molchanov2019importance}
P.~Molchanov, A.~Mallya, S.~Tyree, I.~Frosio, and J.~Kautz, ``Importance
  estimation for neural network pruning,'' in \emph{Proceedings of the IEEE/CVF
  conference on computer vision and pattern recognition}, 2019, pp.
  11\,264--11\,272.

\bibitem{guo2020model}
J.~Guo, W.~Zhang, W.~Ouyang, and D.~Xu, ``Model compression using progressive
  channel pruning,'' \emph{IEEE Transactions on Circuits and Systems for Video
  Technology}, vol.~31, no.~3, pp. 1114--1124, 2020.

\bibitem{jakubovitz2019generalization}
D.~Jakubovitz, R.~Giryes, and M.~R. Rodrigues, ``Generalization error in deep
  learning,'' in \emph{Compressed Sensing and Its Applications: Third
  International MATHEON Conference 2017}.\hskip 1em plus 0.5em minus
  0.4em\relax Springer, 2019, pp. 153--193.

\bibitem{xu2022improving}
W.~Xu, F.~Li, Y.~Jiang, A.~Yong, X.~He, P.~Wang, and J.~Cheng, ``Improving
  extreme low-bit quantization with soft threshold,'' \emph{IEEE Transactions
  on Circuits and Systems for Video Technology}, 2022.

\bibitem{guo2021causal}
Z.~Guo, Z.~Zhang, R.~Feng, and Z.~Chen, ``Causal contextual prediction for
  learned image compression,'' \emph{IEEE Transactions on Circuits and Systems
  for Video Technology}, vol.~32, no.~4, pp. 2329--2341, 2021.

\bibitem{nahshan2021loss}
Y.~Nahshan, B.~Chmiel, C.~Baskin, E.~Zheltonozhskii, R.~Banner, A.~M.
  Bronstein, and A.~Mendelson, ``Loss aware post-training quantization,''
  \emph{Machine Learning}, vol. 110, no. 11-12, pp. 3245--3262, 2021.

\bibitem{flicker2w}
J.~Liu, G.~Lu, Z.~Hu, and D.~Xu, ``A unified end-to-end framework for efficient
  deep image compression,'' \emph{arXiv preprint arXiv:2002.03370}, 2020.

\bibitem{PCGCv2}
J.~Wang, D.~Ding, Z.~Li, and Z.~Ma, ``Multiscale point cloud geometry
  compression,'' in \emph{2021 Data Compression Conference (DCC)}.\hskip 1em
  plus 0.5em minus 0.4em\relax IEEE, 2021, pp. 73--82.

\bibitem{sparse}
C.~Choy, J.~Gwak, and S.~Savarese, ``4d spatio-temporal convnets: Minkowski
  convolutional neural networks,'' in \emph{Proceedings of the IEEE/CVF
  conference on computer vision and pattern recognition}, 2019, pp. 3075--3084.

\bibitem{8iVFB}
E.~d’Eon, B.~Harrison, T.~Myers, and P.~A. Chou, ``8i voxelized full bodies-a
  voxelized point cloud dataset,'' \emph{ISO/IEC JTC1/SC29 Joint WG11/WG1
  (MPEG/JPEG) input document WG11M40059/WG1M74006}, vol.~7, no.~8, p.~11, 2017.

\bibitem{xu2017owlii}
Y.~Xu, Y.~Lu, and Z.~Wen, ``Owlii dynamic human mesh sequence dataset,'' in
  \emph{ISO/IEC JTC1/SC29/WG11 m41658, 120th MPEG Meeting}, vol.~1, 2017, p.~8.

\bibitem{MVUB}
C.~Loop, Q.~Cai, S.~O. Escolano, and P.~A. Chou, ``Microsoft voxelized upper
  bodies-a voxelized point cloud dataset,'' \emph{ISO/IEC JTC1/SC29 Joint
  WG11/WG1 (MPEG/JPEG) input document m38673 M}, vol. 72012, p. 2016, 2016.

\bibitem{hassani2023neighborhood}
A.~Hassani, S.~Walton, J.~Li, S.~Li, and H.~Shi, ``Neighborhood attention
  transformer,'' in \emph{Proceedings of the IEEE/CVF Conference on Computer
  Vision and Pattern Recognition}, 2023, pp. 6185--6194.

\bibitem{wei2022qdrop}
X.~Wei, R.~Gong, Y.~Li, X.~Liu, and F.~Yu, ``Qdrop: Randomly dropping
  quantization for extremely low-bit post-training quantization,'' \emph{arXiv
  preprint arXiv:2203.05740}, 2022.

\bibitem{russakovsky2015imagenet}
O.~Russakovsky, J.~Deng, H.~Su, J.~Krause, S.~Satheesh, S.~Ma, Z.~Huang,
  A.~Karpathy, A.~Khosla, M.~Bernstein \emph{et~al.}, ``Imagenet large scale
  visual recognition challenge,'' \emph{International journal of computer
  vision}, vol. 115, pp. 211--252, 2015.

\end{thebibliography}
\bibliographystyle{IEEEtran}

\clearpage

\appendix

\section*{R-D optimization} 
The quantization-induced performance loss of the underlying task model can be written as Eq. (9), i.e., 
    $ \Delta J \approx \frac{1}{2} E\left[ \left[ \begin{array}{cc}
                    \Delta \boldsymbol{x} \\
                    \Delta \boldsymbol{w}
                \end{array} \right]^T \cdot H_{\boldsymbol{x}, \boldsymbol{w}} \cdot
                \left [\begin{array}{cc}
                    \Delta \boldsymbol{x} \\
                    \Delta \boldsymbol{w} 
                \end{array} \right] \right], $
in the formulation of Sec. IV. We use gradient BP instead of  $ E\left[ \left[ \begin{array}{cc}
                    \Delta \boldsymbol{x} \\
                    \Delta \boldsymbol{w}
                \end{array} \right]^T \cdot H_{\boldsymbol{x}, \boldsymbol{w}} \cdot
                \left [\begin{array}{cc}
                    \Delta \boldsymbol{x} \\
                    \Delta \boldsymbol{w} 
                \end{array} \right] \right]  $   
to optimize $\Delta J$. 

In theory, using
$ E\left[ \left[ \begin{array}{cc}
                    \Delta \boldsymbol{x} \\
                    \Delta \boldsymbol{w}
                \end{array} \right]^T \cdot H_{\boldsymbol{x}, \boldsymbol{w}} \cdot
                \left [\begin{array}{cc}
                    \Delta \boldsymbol{x} \\
                    \Delta \boldsymbol{w} 
                \end{array} \right] \right]  $   
instead of $\Delta J$  as task loss is definitely doable. However, its implementation in practice faces two main challenges:

\begin{enumerate}
\item Hessian matrix $ H_{\boldsymbol{x}, \boldsymbol{w}} $ is a square matrix of second-order partial derivatives of a scalar-valued function or scalar field.  Assuming the dimension of the Hessian matrix is $n$, computing and storing the full Hessian matrix are at an order of $\Theta (n^2) $, i.e., the computation time or required memory increases quadratically with the increase of the input dimension. It is computationally unbearable  for high-dimensional functions such as the loss functions $\Delta J$ used in this paper. 

\item 	Solving $ E\left[ \left[ \begin{array}{cc}
                    \Delta \boldsymbol{x} \\
                    \Delta \boldsymbol{w}
                \end{array} \right]^T \cdot H_{\boldsymbol{x}, \boldsymbol{w}} \cdot
                \left [\begin{array}{cc}
                    \Delta \boldsymbol{x} \\
                    \Delta \boldsymbol{w} 
                \end{array} \right] \right]  $  
optimization is an NP-hard optimization problem. The complexity of solving it scales rapidly with the dimensions of $ \Delta \boldsymbol{x} $ and $\Delta \boldsymbol{w}$. One possible way is to follow AdaRound \cite{nagel2020up} and assume $ H_{\boldsymbol{x}, \boldsymbol{w}} $ is a diagonal matrix, e.g., $ H_{\boldsymbol{x}, \boldsymbol{w}} = c \cdot \boldsymbol{I}$,  by which the optimization objective becomes $\lVert \Delta \boldsymbol{x} / \Delta \boldsymbol{w} \rVert^2$. However, such a simplification way  will lead to a significant loss, as also shown\cite{li2020brecq}. 

\end{enumerate}
Thus, we use gradient BP to optimize $\Delta J$ directly and use 
\begin{align}
    E\left[ \left[ \begin{array}{cc}
                    \Delta \boldsymbol{x} \\
                    \Delta \boldsymbol{w}
                \end{array} \right]^T \cdot H_{\boldsymbol{x}, \boldsymbol{w}} \cdot
                \left [\begin{array}{cc}
                    \Delta \boldsymbol{x} \\
                    \Delta \boldsymbol{w} 
                \end{array} \right] \right] \label{eq:simp_task_loss_response}
\end{align}
to describe localized non-monotonic behavior for the motivation of R-D loss optimized PTQ. 

\section*{Point Cloud Compression Task} 
As examined in the paper, we show that task-loss-optimized PTQ is a generic and promising solution for learning-based model quantization, with examples of RDO-PTQ for learned image compression, distortion-loss-optimized PTQ for super-resolution, and cross-entropy loss-optimized PTQ for classification. 

As suggested by the reviewer, we further extend the idea to 3D point cloud geometry compression. With this aim,  we select the PCGCv2 \cite{PCGCv2} as our baseline because of its wide recognition. PCGCv2 applies  sparse convolutions to characterize and embed information upon valid elements in sparse tensor~\cite{sparse}, which is significantly different from 2D image processing with well-structured pixel grids. We adopt the same settings used in experiments in the original floating-point PCGCv2.  Three different datasets, e.g.,  8i Voxelized Full Bodies (8iVFB) \cite{8iVFB}, Owlii dynamic human meshh \cite{xu2017owlii}, and  Microsoft Voxelized Upper Bodies (MVUB) \cite{MVUB} are examined, and seven bitrates are applied for each sample sequence. The bit rate is measured using bits per input point (bpp), and the distortion is evaluated using point-to-point distance (D1) and point-to-plane distance (D2) based mean squared error (MSE).  We calculate the BD-rate loss over the original models in Table~\ref{tab:PCGCv2_response}

\begin{table}[htbp]
  \centering
  \caption{ BD-Rate loss over the floating-point PCGCv2~\cite{sparse}.  Both  D1 and D2 distortion measurements are evaluated.}
    \begin{tabular}{ccccccc}
    \toprule
          & \multicolumn{2}{c}{bit-width} & \multicolumn{4}{c}{Datasets} \\
    Distortion & w & a & 8iVFB & Owlii & MVUB  & Average \\
    \midrule
    D1    & 6     & 6     & 1.77\% & 2.73\% & 1.35\% & 1.95\% \\
    D2    & 6     & 6     & 1.42\% & 2.17\% & 2.38\% & 1.99\% \\
    \bottomrule
    \end{tabular}%
  \label{tab:PCGCv2_response}%
\end{table}%

As shown in Table.~\ref{tab:PCGCv2_response}, even at  6-bit precision, we achieve only less than $2\%$ average BD-rate loss after model  quantization, further suggesting that the proposed task-loss-optimized PTQ methodology is generalized.


\end{document}